\documentclass[11pt]{article}
\pdfoutput=1

\usepackage{epsfig,latexsym,amsfonts,amsmath,amsthm,amssymb,amsbsy,multirow,slashed,wasysym,textcomp,dsfont,comment,mathtools,cancel,cite,diagbox,datetime,appendix}
\usepackage{bbm}
\usepackage{tocloft}
\usepackage{sidecap}
\usepackage{adjustbox}
\usepackage{pgfplots}
\usepackage{oplotsymbl}
\usepackage{tikzpagenodes}
\usepackage{adforn}
\usepackage{tikz}
\usetikzlibrary{backgrounds}
\usetikzlibrary{patterns}
\usetikzlibrary{arrows.meta}
\usetikzlibrary{shapes,decorations}
\tikzstyle{bag} = [align=center]
\usetikzlibrary{decorations.pathmorphing}
\usepackage[normalem]{ulem}
\usepackage[mathscr]{eucal}
\usepackage{stackengine}
\stackMath
\usepackage{hyperref}
\usepackage{subcaption}
\usepackage{tensor}
\usepackage{physics} 

 \newcommand{\badat}{\begin{alignedat}}
 \newcommand{\eadat}{\end{alignedat}}
 
 \def\be{\begin{equation}}
\def\ee{\end{equation}}

\usepackage{color}
\newcommand{\red}[1]{\textcolor{red}{#1}}

\newcommand{\blue}[1]{\textcolor{blue}{#1}}

\newcommand{\pink}[1]{\textcolor{\pink}{#1}}

\definecolor{dred}{rgb}{0.65,0.10,0.20} 
\definecolor{dblue}{rgb}{0.2,0.50,0.80}

\def\Tr{{\rm Tr}}

  \newcommand{\coo}[1]{{\mathcal{O}\left(#1\right)}}
    \newcommand{\avg}[1]{{\left\langle#1\right\rangle}}
	\def\aJ{{|J|}}
    
	\def\cD{{\cal D}}
 \def\cN{{\cal N}}

    \def\pd{{\partial}}
    \def\rb{{\bar{\rho}}}

\numberwithin{equation}{section} 
\pgfplotsset{compat=1.17}

\hypersetup{
  colorlinks   = true, 
  linkcolor={blue!50!black},
    citecolor={red!70!black},
    urlcolor={blue!80!black}
}

\begin{document}


\begin{titlepage}
  \thispagestyle{empty}

\begin{center}

 {\LARGE \scshape{Size and Shape of Rotating Strings \\[10pt] and the Correspondence to Black Holes
}}

 \vskip1cm 

   \centerline{ 
   {Nejc \v{C}eplak}${}^\spadesuit{}$,
   {Roberto Emparan}${}^{\heartsuit{}_{a,b}}$,
   {Andrea Puhm}${}^\blacklozenge{}$,
   {and Marija Tomašević}${}^\blacklozenge{}$
   }


\bigskip\bigskip
 
\noindent{\em${}^\spadesuit$
School of Mathematics and Hamilton Mathematics Institute, \\Trinity College, Dublin 2, Ireland
}
 
\bigskip
 
\noindent{\em${}^{\heartsuit{}_{a}}$
Institució Catalana de Recerca i Estudis Avançats (ICREA)\\
Passeig Lluis Companys, 23, 08010 Barcelona, Spain\\
${}^{\heartsuit{}_{b}}$ Departament de Física Quàntica i Astrofísica and
 Institut de Ciències del Cosmos,\\
 Universitat de Barcelona, 08028 Barcelona, Spain
 }
 
\bigskip
 
\noindent{\em${}^\blacklozenge$ 
Institute for Theoretical Physics, University of Amsterdam,\\ 
PO Box 94485, 1090 GL Amsterdam, The Netherlands
}

\end{center}

\medskip
%

\centerline{\footnotesize\upshape\ttfamily  ceplakn@tcd.ie, emparan@ub.edu, a.puhm@uva.nl, m.tomasevic@uva.nl
 }

\bigskip

\begin{abstract}
  \noindent 
 In light of the correspondence between black holes and fundamental strings with non-zero spin, we compute the sizes of rotating strings for small, moderate, and large values of the angular momentum and compare them to the sizes of rotating black holes.
 We argue that the ratio of the size perpendicular to the rotation plane to the size along the rotation plane is an approximate adiabatic invariant. and can therefore be meaningfully compared for objects on different sides of the correspondence point.
 We show that the spin-dependence of this ratio for small angular momenta agrees for black holes and strings. When the spin is large, the ratios for these objects exhibit different behavior, but
 this is expected since for large angular momenta there is no direct correspondence between black holes and single-string states. 
 We also develop a random-walk model that describes highly excited strings and accurately reproduces the sizes of rotating strings for all values of angular momentum.
 
\end{abstract}

\end{titlepage}

\setcounter{tocdepth}{2}
{\small \tableofcontents}

\newpage

\section{Introduction and summary}
\label{sec:Intro}

The massive states of a fundamental string share with black holes the property of being large and highly degenerate objects. This similarity has been taken further to propose a smooth transition between them when the `t~Hooft-like coupling $g^2 S$ is of order one (with $g$ being the closed string coupling and $S$ the entropy of the state) \cite{Susskind:1993ws,Horowitz:1996nw,Chen:2021dsw}. Recently we have examined this transition for states with non-zero angular momentum \cite{Ceplak:2023afb}. In this article we extend the study to compute the size and shape of highly excited spinning strings and then compare them to black holes.

\paragraph{Sizing up spinning black holes and strings.}
The first consequence of adding spin to both objects is to shrink their overall size. For a black hole, this can be seen in at least two contrasting ways.
The centrifugal effect of rotation (or, more generally, the gravitational repulsion created by angular momentum) counteracts the gravitational attraction of mass, making the region that is trapped smaller and thus reducing the size of the event horizon.\footnote{The shrinkage when charging a black hole can be explained with similar arguments.}  A complementary perspective is afforded if we regard the mass $M$ of the black hole as consisting of two components: heat, $T S$, and mechanical energy, $\Omega J$. If we increase the rotation, then the heat fraction for a given total energy must be smaller. Since the heat, i.e., the entropy of a black hole, has a geometrical counterpart as horizon area, we conclude that the overall size of a black hole of mass $M$ must be smaller for larger $J$. The general relation
\begin{align}\label{MPD}
   \left(\frac{M}{M_P}\right)^{D-2} \propto S^{D-5}\left( \frac{S^2}{4\pi^2} + J^2\right)\,
\end{align}
(with $M_P$ the Planck mass), which is valid for Myers-Perry black holes in $D\geq 4$ \cite{Myers:1986un}, bears this out.

This second kind of argument also applies, with a small variation, to the states of a free string. The oscillator excitations in a typical massive state can be divided into two classes: oscillation modes that make the string coherently rotate, and oscillations that make the string jitter in a random-walk fashion and give the large degeneracy of states. If more oscillator energy is used for rotation, then less is available to random-walk. Since the size of a random walk grows with the (square-root of the) number of random steps, we conclude again that increasing the total angular momentum should reduce the overall size of the string ball. This is apparent in the formula for the entropy of a rotating closed string with very large spin, $|J|=\coo{M/M_s}^2$,
\begin{align}\label{Sstr}
    S^2\propto \left(\frac{M}{M_s}\right)^2-2|J|\,,
\end{align}
where $M_s=1/\sqrt{\alpha'}$ is the string mass scale \cite{Russo:1994ev}. We will confirm that this measures the random spread of a typical rotating string in this regime of large spins.

Although \eqref{MPD} and \eqref{Sstr} similarly express the qualitative effect that rotation has on overall size, they are obviously rather different formulas. Then, the possible correspondence between the two kinds of objects may look, at the very least, very imprecise as to the way that spin affects the states.
In this article we show that things are better than this observation suggests. We extend and refine the study of the properties of rotating string states to include more detailed information about their shape in different directions, and in regimes from small spins, $J=\order{1}$, to the largest ones closer to the Regge bound, $|J|=M^2/(2 M_s^2)$. This is necessary, since in \cite{Ceplak:2023afb} we explained that, even though string states and black holes can exist with ultraspins $|J|>S$ (in $D\geq 5$), the black hole/string correspondence only matches them in regimes of moderate spins $|J|\lesssim S$. Since \eqref{Sstr} is derived well outside this regime, it is not surprising that it does not bear much resemblance to \eqref{MPD}.

In particular, we derive the string density of states (and thus the entropy) and calculate the average sizes of strings parallel and orthogonal to the rotation plane, $r_\|$ and $r_\perp$, for small spins, spins close to the Regge bound $J\sim \coo{M/M_s}^2$, and for 
string states in an intermediate regime where the angular momentum scales as $J\sim \coo{M/M_s}$.
These sizes are expected to vary when the string coupling $g$ grows and self-gravitation effects become important, i.e., they are not approximately adiabatic invariants that would vary little between the free string and the strongly gravitating black hole regimes\footnote{As emphasized in \cite{Ceplak:2023afb}, for non-BPS states the black hole/string correspondence can only involve an approximate notion of `Goldilocks adiabaticity' since at any finite and non-zero coupling the states involved are never strictly stationary.}. For a string, the sizes are naturally measured in units of the string length $\ell_s$, while for a black hole they are measured in units of the Planck length 
\begin{align}
    \ell_P=g^{\frac{2}{D-2}}\ell_s\,.
\end{align}
Even taking this rescaling into account, the effects of self-interaction on the string ball are required to match its size to that of a black hole at the correspondence point \cite{Horowitz:1997jc,Damour:1999aw,Chen:2021dsw}. Self-interaction makes the string ball shrink overall, but in a highly dimension-dependent manner, and in this article we do not study this effect for rotating balls (in this direction, see \cite{Santos:2024ycg}).

\paragraph{Spin dependence.}
While $r_\|$ and $r_\perp$ vary significantly with the strength of interaction, it seems reasonable to expect that they are similarly affected. Then, the ratio $r_\perp/r_\|$ has a weaker dependence on $g$, at least for moderate values, and behaves as an approximate adiabatic invariant.\footnote{$J$ and $S$ are adiabatic invariants, while the mass $M$ and angular velocity $\Omega$ will, in general, vary significantly with the strength of self-interaction.} 
It is natural that $r_\perp/r_\|$ decreases as $|J|$ grows, indicating that rotation makes the object spread in the rotation plane. We verify this effect and show that when the spin $J$ is much smaller than $S$, both black holes and string states satisfy
\begin{align}\label{perptopar}
    \frac{r_\perp^2}{r_\|^2}-1 \propto -\frac{J^2}{S^2}\,,
\end{align}
where the proportionality factor is an $\order{1}$ number that, for several reasons that we detail below, we cannot expect to exactly match between the two sides. 
Although $J^2/S^2$ for $|J|\ll S$ is a small number, the functional dependence is significant~---~unlike other $\order{1}$ numerical values to which the correspondence is insensitive.
For a black hole, the dependence $\sim J^2/S^2$ is evident from \eqref{MPD}, but for a string state it is less obvious 
and is derived here for the first time. A generic dependence $\propto J^2$ for string states may be inferred from symmetry under $J\to -J$ and the assumption of analyticity as $J\to 0$, but the precise form $J^2/S^2$ involves subtler considerations.

To examine this point further, let us temporarily restore $\hbar\neq 1$. For a semiclassical black hole with a finite horizon area, the entropy scales as $S\propto \hbar^{-1}$ and is a large number. To ensure the angular momentum appears as a semiclassical quantity, 
$J$ must also be large in units of $\hbar$. Thus we see that the dimensionless ratio $J/(\hbar S)$ is a quantity that remains finite in the semiclassical limit. This (and the symmetry $J\to -J$) is the reason why the leading spin dependence in the semiclassical black hole takes the form $J^2/S^2$. Observe now that a term $J^2/S^3$, which as a dimensionless quantity corresponds to $\hbar\, J^2/(\hbar S)^3$, is a quantum correction $\propto \hbar$ in the semiclassical expansion. Therefore, we do not expect to find it for black holes unless we account for one-loop effects. In contrast, a term of the form $J^2/S$, which corresponds to $\hbar^{-1} J^2/(\hbar S)$, cannot appear in the semiclassical expansion since it would diverge when $\hbar \to 0$ (unless somehow $S\sim \hbar^{-2}$, which we find impossible for a black hole).

For fundamental strings at level $n$, the entropy scales as $S\propto \sqrt{n}$ (with no factor of $\hbar$), and we focus on states with $n\gg 1$. In the spin expansion, terms $\propto J^2/n$ naturally correspond to the leading terms $\propto J^2/S^2$ for semiclassical black holes, and indeed we find them in the size ratio. Actually, the first terms in the spin expansion of the density of states and the sizes are integer powers of $J^2/n$. Similarly, in the string calculations we find terms $\propto J^2/n^{3/2}$ which, as we have argued, correspond to quantum corrections on the black hole side (which are expected but we have not computed). However, a term $\propto J^2/n^{1/2}$ cannot have a black hole counterpart, and indeed no such terms arise in our analysis of rotating strings. 
To be sure, their absence is not due to any miraculous cancellation; a detailed examination of the structure of the spin expansion calculation shows that such terms are bound to be absent in the final result. This would not be a priori obvious without an explicit analysis, which not only confirms the absence of $J^2/n^{1/2}$, but also establishes the presence of $J^2/n$ as the leading-order contributions. If the latter had not happened, it would have been equally problematic. Our study thus demonstrates that the microstates of fundamental strings carry spin in a way that, in the limit of very large masses, can be put in correspondence to semiclassical black holes -- a conclusion that we regard as significant.

\subsection*{Summary of our results}

The key results of our analysis of sizes of rotating strings using both an operator method and a random walk model and their comparison with black holes are as follows.

\begin{figure}[!ht]
    \centering
    \includegraphics[width = 0.9\linewidth]{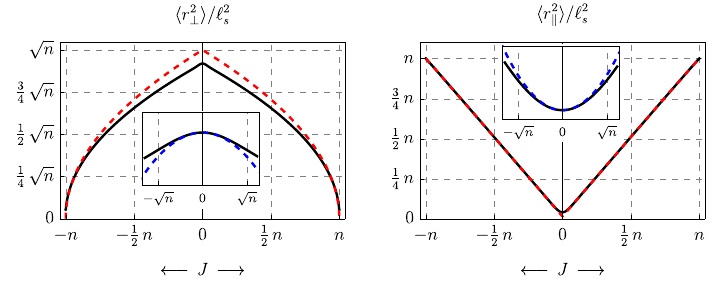}
    \caption{\small Sizes of strings in string units in the directions orthogonal to the plane of rotation (left) and in the plane of rotation (right). 
    In both plots, the {\bf black} curve represents the size obtained by numerically solving the saddle point equation. The  {\color{red} \bf red} dashed line represents the analytic result in the regime where $J \sim \coo{n}$. In the inset, we zoom in to the regime of $J\sim \coo{1}$, where we show with the {\color{blue} \bf blue} dashed line the agreement with the quadratic corrections analysed in \eqref{eq:ResPerp} and \eqref{eq:ResParallel}. 
    For the purposes of these illustrations, we have ignored any constant numerical prefactors. }
    \label{fig:Size1}
\end{figure}

\subparagraph{Sizes of strings.} For string states at excitation level $n\gg 1$ and spin $J$ the average sizes, in string units, in directions
transverse to the plane of rotation are (see~\eqref{rperp} and \eqref{rperpsmallJ})
\begin{equation}
\label{eq:ResPerp}
      \frac{\langle \bar{r}^2_\perp \rangle_n}{\ell_s^2}\propto 
\begin{cases}
     \sqrt{n} +\gamma_1+\frac{\gamma_2}{\sqrt{n}}
     , &  J = \order{1} \\
    \sqrt{n-\mu_d|J|} , &  J = \order{\sqrt{n}} \\
    \sqrt{n-|J|} , & J = \order{n}
    \end{cases},
\end{equation}
while along the plane of rotation they are given by (see~\eqref{rparallel} and~\eqref{rparallelsmallJ})
\begin{equation}
\label{eq:ResParallel}
    \frac{\langle \bar{r}^2_\parallel \rangle_n}{\ell_s^2} \propto \begin{cases}
     \sqrt{n} +\gamma_1+\frac{\gamma_{2}}{\sqrt{n}} + \frac{\gamma_{\parallel}}{\sqrt{n}}  J^2
     , &  J = \order{1} \\
    \sqrt{n-\mu_d|J|} \,C_\parallel, &  J = \order{\sqrt{n}} \\
     |J| , & J = \order{n}
    \end{cases}.
\end{equation}
Here  $\mu_d=\tanh({\beta|J|/2})\in(0,1)$, the positive constants $\gamma_1$, $\gamma_2$ and $ \gamma_\parallel$ are given in~\eqref{eq:consts}, and $C_\parallel$ is a slowly varying function of $|J|/\sqrt{n}\sim \order{1}$, given in \eqref{Ceq}, and whose most relevant property is that $C_\parallel>1$. 
These results are depicted in Figure~\ref{fig:Size1}.

The approximate adiabatic invariant ratio between the average sizes transverse to and along the rotation plane is
\begin{equation}
\label{eq:ResRatio}
    \frac{\langle \bar{r}^2_\perp \rangle_n}{\langle \bar{r}^2_\parallel \rangle_n} \propto \begin{cases}
    1-\gamma_\parallel \frac{J^2}{S^2}
     , &  J = \order{1} \\
   \frac{1}{C_\parallel}<1, &  J = \order{\sqrt{n}} \\
      \frac{S}{|J|} , & J = \order{n}
    \end{cases}.
\end{equation}
Observe that for small spins this is of the form \eqref{perptopar}.

\subparagraph{Sizes of black holes.}
For Myers-Perry black holes the definitions of the transverse and parallel sizes are not completely unambiguous due to the considerable curvature of the geometry, but when we make choices that are both natural and simple (see Section~\ref{sec:BHSize}), then we find
\begin{equation}\label{eq:bhratios}
       \frac{r_\perp^2}{r_\|^2}=
\begin{cases}
    1-4\pi^2 \frac{J^2}{S^2}, &  |J| \ll S \\ \text{const.}<1, &  |J| \sim S \\
    \frac{1}{4\pi^2}\frac{S^2}{J^2} , & |J| \gg S
    \end{cases}\,.
\end{equation}
These are valid in any dimension $D> 4$, and in $D=4$ for $2\pi|J|\leq S$. Other black objects in $D\geq 5$, like black rings and black bars, only exist with $|J|>S$. For $|J| \gg S$ they also satisfy $r_\perp^2/r_\|^2\propto S^2/J^2$.

\begin{figure}[ht!]
    \centering
    \includegraphics[width=0.9\linewidth]{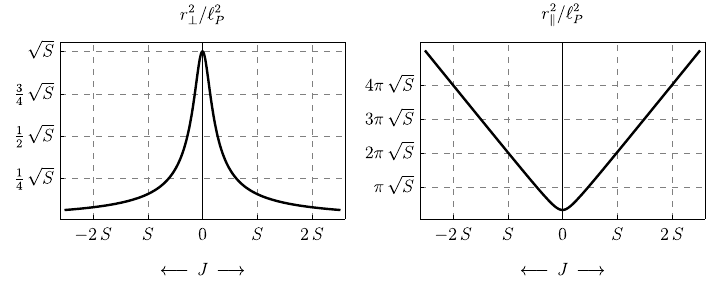}
    \caption{\small Sizes of black holes in $D=6$ in appropriately chosen Planck units.%
    }
    \label{fig:bhsize}
\end{figure}

\begin{figure}[ht!]
    \centering
    \includegraphics[width=0.9\linewidth]{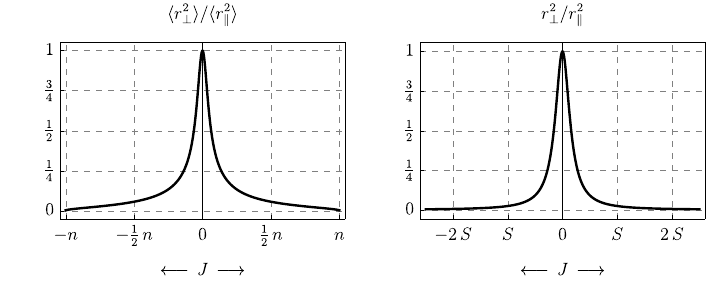}
    \caption{\small Ratios of the sizes for the strings (left) and black holes (right). Since these ratios are approximate adiabatic invariants, the resemblance between the graphs is significant, especially at small and moderate $J$.}
    \label{fig:ratios}
\end{figure}

\subparagraph{Comparison.} The sizes $r_\perp$ and $r_\|$ differ between slowly spinning  strings and black holes with $J\sim \order{1}$ and $S\sim \sqrt{n}$ since we are ignoring the self-gravitation of the strings. Figure~\ref{fig:Size1} and Figure~\ref{fig:bhsize} show these sizes, and although the parallel sizes $r_\|^2$ exhibit a similar dependence on the spin, the transverse sizes $r_\perp^2$ differ more markedly.
However, for both objects the ratio $r_\perp/r_\|$ in this regime
reproduces \eqref{perptopar} in agreement with the expected correspondence. The plots for these ratios in Figure~\ref{fig:ratios} bear a remarkable resemblance up to rather large spins.

For $|J|>S$ not only the separate values, but also the ratio of perpendicular and parallel sizes begins to differ between the two objects. However, this is not unexpected, since we argued in \cite{Ceplak:2023afb} that, at large spins, single-string states do not correspond to stationary rotating black holes.

\subparagraph{Rotating strings as random walks.}
Highly excited strings heuristically behave like random walks \cite{Salomonson:1985eq, Mitchell:1987hr, Mitchell:1987th, Horowitz:1997jc,Abel:1999rq,Barbon:2004dd, Manes:2004nd, Kruczenski:2005pj}. 
We aim to make this picture more precise by introducing a random walk model for rotating strings.
Using this model we reproduce the sizes of quantized rotating strings in \eqref{eq:ResPerp} and \eqref{eq:ResParallel} for all values of the angular momentum $J$ and provide an alternative geometric interpretation of these results.

Our analysis shows that our random walk model contains information about the spatial distribution of highly excited strings and provides an accurate description of them even for arbitrary values of the angular momentum. We expect that it can be used to extract properties beyond their sizes, such as higher multipole moments.

\bigskip

The plan for the rest of the paper is the following. In Section~\ref{sec:sizes}, we compute the sizes of rotating strings for all relevant regimes of angular momenta, and in Section~\ref{sec:BHSize}, we do the same for rotating black holes of all shapes and dimensions. In Section~\ref{sec:randomwalk}, we introduce a random walk model for rotating strings and show that it reproduces the sizes of strings for all values of the angular momentum.
The appendices contain computational details omitted in the main text.
In Appendix~\ref{app:rhotation}, we extract from the infinite sum that appears in the expression for the size of strings in the plane of rotation a simple expression that characterizes the effect of rotation in the high-temperature limit.
Appendix~\ref{app:Fourier} and~\ref{app:Sizes} contain details of how to evaluate the integral transforms that trade the inverse temperature~$\beta$ and angular velocity~$\Omega$, respectively, for the excitation level~$n$ and angular momentum~$J$, and which thus play a key role in determining the sizes of strings as a function of $n$ and $J$. 
In Appendix~\ref{sec:detsrot}, we repeat the computation of sizes for closed fundamental strings.
In Appendix~\ref{app:RandomWalk} we lay out the details for the random walk model for rotating strings.
In Appendix~\ref{sapp:CanEns} we discuss the string sizes in terms of $\beta$ and $\Omega$ and show how to obtain the scalings in \eqref{eq:ResPerp} and \eqref{eq:ResParallel} using thermodynamic relations, bypassing the evaluation of the integral transforms that trade $\beta$ for $n$ and $\Omega$ for $J$.
%

\section{Size and shape of strings}
\label{sec:sizes}

In this section we derive the sizes of rotating strings for small, moderate, and large values of angular momenta, respectively $J=\order{1},\, \order{\sqrt{n}},\, \order{n}$, with $n$ the oscillator level. We first review the static calculation of \cite{Mitchell:1987hr}, and then apply a similar operator method to extract the parallel and transverse sizes of rotating strings. We are interested in the ratio of these two sizes since, as we have explained, it can be seen as an approximate adiabatic invariant.
\subsection{Review: static strings}
\label{sec:StaticString}

Consider a closed bosonic string in $D$ spacetime dimensions,
\begin{equation}%
\label{Xclosed0}
    X^\mu(\tau,\sigma)=x_{\rm c.o.m.}^\mu+\alpha'\ p^\mu \tau +i\sqrt{\frac{\alpha'}{2}} \sum_{n=-\infty,n\neq 0}^\infty \frac{1}{n}\left(\alpha^\mu_n e^{-in(\tau+\sigma)}+\tilde \alpha^\mu_n e^{-in(\tau-\sigma)}\right),
\end{equation}
where $\tau$ and $\sigma$ denote the world-sheet coordinates.
We then work in the light-cone gauge, defining $X^\pm =\frac{1}{\sqrt{2}}(X^0\pm X^{D-1})$ and denoting the $D-2$ transverse coordinates $X^i$.
The average size of a static closed string can be estimated as follows. Singling out one of the transverse directions, $X^i$,  we consider the position operator  evaluated at $\tau=0$, \begin{equation}\label{Xclosedi}
    X^i(\tau=0,\sigma)=x_{\rm c.o.m.}^i +i\sqrt{\frac{\alpha'}{2}} \sum_{n=-\infty,n\neq 0}^\infty \frac{1}{n}\left(\alpha^i_n e^{-in\sigma}+\tilde \alpha^i_n e^{+in\sigma}\right)\,.
\end{equation}
Keeping in mind that we have rotational symmetry in the $D-1$ spatial directions, the mean squared radius of the string is given by
\begin{equation}\label{Deltar}
\bar r^2=\frac{D-1}{\pi} \int_0^\pi d\sigma : (X^i(0,\sigma)-x^i)^2\equiv\alpha'(D-1) (R^2+\tilde R^2)+\alpha \tilde \alpha\; {\rm terms}\,,
\end{equation}
where $:\, :$ denotes normal ordering and we defined
\begin{equation}\label{Rsize}
    R^2\equiv\sum_{n=1}^\infty\frac{1}{n^2} \alpha^i_{-n} \alpha^i_n\,,
\end{equation}
and a similar expression for $\tilde R^2$.
We are interested in the expectation value of $\bar r^2$ averaged over all states at level~$n$ (to which the $\alpha \tilde \alpha$ terms do not contribute). Focusing on the contribution from $R$, this amounts to computing
\begin{equation}\label{rbarn}
    \langle \bar r^2\rangle_n \propto \frac{R^2_n}{d_n},
\end{equation}
where the density $d_n$ of states at level $n$ and $R^2_n$ are extracted from the expansions of the partition function and the average size squared,
\begin{equation}\label{R}
   Z(x)\equiv \Tr(x^N)=\sum_{n=1}^\infty d_n x^n\,,\qquad 
   \langle R^2 \rangle \equiv
   \Tr(R^2 x^N)=\sum_{n=1}^\infty R^2_n x^n,
\end{equation}
where 
\begin{equation}\label{eq:xbeta}
    x\equiv e^{-\beta}
\end{equation}
and $\beta$ is the inverse temperature.

\subsubsection{Density of states}
\label{sec:StaticStringDensity}
%

The level-counting operator is given by
\begin{equation}
\label{eq:LevelCount}
    N=\sum_{n=1}^\infty \sum_{i=1}^{c} \alpha_{-n}^i \alpha_n^i,
\end{equation}
where henceforth we denote the number of transverse oscillators by 
\begin{equation}\label{eq:defc}
    c\equiv D-2\,.
\end{equation}
The partition function
\begin{equation}
Z(x)=
\prod_{n=1}^{\infty}\left(\frac{1}{1-x^n}\right)^{c}
\end{equation}
can be expressed in terms of Jacobi theta and Dedekind eta functions. We can use their modular transformation properties to compute the high-temperature behavior.  As  $\beta\to 0$ this yields
\begin{equation}
\label{Zbetathermo}
    Z(\beta \approx 0) =\text{const.}\, \beta^{\frac{c}{2}} e^{\frac{c \pi^2}{6 \beta}}\,,
\end{equation}
up to a numerical factor. 
We can extract the density of states with level $n$, $d_n$, from the expansion
\begin{equation}
    Z(x)=\sum_{n=1}^\infty d_n x^n\,,\qquad d_n=\frac{1}{2\pi i} \oint \frac{dx}{x^{n+1}} Z(x)\,,
\end{equation}
via Cauchy's theorem taking $x$ complex.
In the high-temperature limit, the density 
\begin{equation}
\label{eq:StaticDensity}
    d_n=-\frac{1}{2\pi i} \int d\beta e^{n\beta+\log Z(\beta)}
    \equiv -\frac{1}{2\pi i}\int d\beta e^{nf(\beta))},
\end{equation}
where
\begin{equation}
    n f(\beta)=\frac{a}{\beta}+b\log \beta+\beta n\,,
\end{equation}
with 
\begin{equation}\label{eq:defab}
    a=\frac{c \pi^2}{6}\,, \qquad b=\frac{c}{2} \,,
\end{equation}
can be obtained using the steepest descent method. To this effect, we pick a contour that goes through the saddle point 
\begin{equation}\label{staticsaddle}
    \beta_*=\sqrt{\frac{a}{n}}+\order{n^{-1}},
\end{equation}
where $f'(\beta_*)=0$.
Up to a numerical factor, the density of states of high-energy strings is given by
\begin{equation}\label{dnTM}
 d_n =\text{const.}\, \left(n |f''(\beta_*)| \right)^{-1/2} e^{n f(\beta_*)}=\text{const.} \,n^{-\frac{c+3}{4}} e^{2\pi \sqrt{c n/6}}\,.
\end{equation}
This is of course the classic Hardy-Ramanujan result \cite{10.1112/plms/s2-17.1.75, Green:1987sp} for $c$ oscillators.

\subsubsection{Size and shape}
\label{sec:StaticStringSize}

The average size squared can be computed from
\begin{equation}
\label{R2static}
    \langle R^2 \rangle \equiv\Tr (R^2 x^N) =\frac{d}{dy}\Tr ( x^N y^{R^2})\Big|_{y=1}\,.
\end{equation}
Because of rotational symmetry in the $D-1$ spatial directions, we can choose any direction in~\eqref{Rsize}. The trace 
\begin{equation}
 \Tr (x^N y^{R^2})=\prod_{n=1}^\infty  \left(\frac{1}{1-x^n}\right)^{D-3}  \frac{1}{1-x^ny^{1/n}}\,,
\end{equation}
then extracts the size of the string ball. The average size squared is 
\begin{equation}
    \langle R^2 \rangle =\prod_{n=1}^\infty \left(\frac{1}{1-x^n}\right)^{D-2}  \sum_{m=1}^\infty \frac{x^m}{m(1-x^m)}.
\end{equation}
To extract the coefficient $R^2_n$ in \eqref{R} in the thermodynamic limit 
note that
\begin{equation}
\label{statsize}
     \langle R^2\rangle=\frac{1}{c} Z(\beta) \log Z(\beta).
\end{equation}
Thus the computation of $R^2_n$ follows the same steps as the one for the density of states, with the only difference being the extra factor 
\begin{equation}
    \frac{1}{c}\log Z(\beta)\stackrel{\beta\to 0}{\approx} \frac{1}{c}\frac{a}{\beta}+\frac{1}{2}\log \beta\,.
\end{equation}
Evaluating $\beta$  at the saddle point \eqref{staticsaddle} yields, to leading order,
\begin{equation}
    R^2_n=\frac{1}{c}\sqrt{a n}\,d_n.
\end{equation}
For a typical configuration of the string, we have that its mass is $M\propto \sqrt{n}$, and so 
the size 
\begin{equation}
\label{eq:StaticSize}
    \langle\bar r^2\rangle_n\propto \sqrt{n} \propto M
\end{equation}
agrees, as expected, with that of a random walk \cite{Manes:2004nd}.

\subsection{Rotating strings}
\label{sec:RotatingStringSize}

We now introduce angular momentum following \cite{Russo:1994ev}. We consider a bosonic open string with mass $M$ and with angular momentum $J$ carried by a subset of its oscillators. The effect of rotation is most pronounced when restricting to a single rotation plane.
The worldsheet Hamiltonian is 
\begin{equation}
\label{eq:Hamil1}
    H=N-\Omega J=\sum_{n=1}^\infty \sum_{i=1}^{D-2} \alpha_{-n}^i \alpha_n^i -\Omega J,
\end{equation}
and we single out the $i=1,2$ oscillators such that
\begin{equation}
\label{eq:AngularMom}
 J=-i\sum_{n=1}^\infty \frac{1}{n} (\alpha_{-n}^1 \alpha_n^2-\alpha_{-n}^2 \alpha_n^1)
\end{equation}
generates rotation in the 12--plane and $\Omega$ is a Lagrange multiplier corresponding to the angular velocity of the string in the rotation plane. We will also consider closed strings where the Hamiltonian is \eqref{eq:Hamil1}  with $J \to J_L + J_R$; the details of the computation for the density of states and sizes are relegated to Appendix~\ref{sec:detsrot} and we summarize the results at the end of this section.

We can diagonalize the Hamiltonian \eqref{eq:Hamil1} by introducing operators $a,b,a^\dagger,b^\dagger$ such that
\begin{equation}
    \alpha_n^1=\sqrt{\frac{n}{2}}(a_n+b_n), \quad \alpha_n^2=-i\sqrt{\frac{n}{2}} (a_n-b_n),\quad
    \alpha_{-n}^1=\sqrt{\frac{n}{2}}(a_n^\dagger+b_n^\dagger), \quad   \alpha_{-n}^2=i\sqrt{\frac{n}{2}}(a_n^\dagger-b_n^\dagger),
\end{equation}
whose commutators are $[a_n,a_m^\dagger]=\delta_{nm}$, $[b_n,b_m^\dagger]=\delta_{nm}$. In terms of these
\begin{equation}
    J=-\sum_{n=1}^\infty  (a_{n}^\dagger a_n-b_n^\dagger b_n)
\end{equation}
and the Hamiltonian takes the form
\begin{equation}
    H=\sum_{n=1}^\infty \left( \sum_{i=3}^{D-2} \alpha_{-n}^i \alpha_n^i +(n+\Omega) a_n^\dagger a_n +(n-\Omega) b_n^\dagger b_n\right).
\end{equation}

For the rotating string, we can define the size parallel and perpendicular to the rotation plane,
\begin{equation}
        R_\parallel^2\equiv\sum_{n=1}^2\frac{1}{n^2} (\alpha^i_{-n}\alpha^i_n),\qquad  R_\perp^2\equiv\sum_{n=3}^\infty\frac{1}{n^2} (\alpha^i_{-n}\alpha^i_n),
\end{equation} 
and similar expressions for $\tilde R_\parallel^2$ and $\tilde R_\perp^2$. We are interested in the expectation values 
\begin{equation}
\label{rsizes}
    \langle \bar r_\parallel^2\rangle_n \propto \frac{R_{n,\parallel}^2}{d_{n,J}}\,, \qquad \langle \bar r_\perp^2\rangle_n \propto \frac{R_{n,\perp}^2}{d_{n,J}}\,,
\end{equation}
where $R_{n,\parallel}^2$ and $R_{n,\perp}^2$ are extracted from the expansions
\begin{equation}\label{Rparallel_Rperp}
    \langle R_\parallel^2\rangle\equiv\Tr(R_\parallel^2\, x^{N-\Omega J})=\sum_{n=1}^\infty R_{n,\parallel}^2 \,x^{n-\Omega J}, \qquad \langle R_\perp^2\rangle\equiv\Tr(R_\perp^2\, x^{N-\Omega J})=\sum_{n=1}^\infty R_{n,\perp}^2\, x^{n-\Omega J}, 
\end{equation}
(with $x$ as in \eqref{eq:xbeta}) while the density $d_{n,J}$ of states at level $n$ and spin $J$ is extracted from the partition function for the rotating string
\begin{equation}
\label{eq:PartFunDec}
\Tr(x^{N-\Omega J})=\sum_{n,J}d_{n,J} e^{-\beta(n- \Omega J)}\,.
\end{equation}

In what follows, we extract $d_{n,J}$ using two integral transforms, similar to what was done in the static case \eqref{eq:StaticDensity}.
When one is interested only in the scaling of the density of states and sizes of rotating strings with $n$ and $J$, one can bypass the evaluation of the two integrals by using thermodynamic relations between pairs of conjugate variables ($\beta$ and $n$, $\Omega$ and $J$). 
We discuss this method in more detail in Appendix~\ref{sapp:CanEns}.

\subsubsection{Density of states}
\label{sec:RotatingStringDensity}
 
The partition function for a string rotating in one plane is given by
\begin{equation}
\label{ZxOm}
  Z(x,\Omega)\equiv \Tr(x^{N-\Omega J})=\prod_{n=1}^\infty \left(\frac{1}{1-x^n}\right)^{D-4}\frac{1}{(1- x^{n+\Omega})(1-x^{n-\Omega})}\,.
\end{equation}
We want to extract the density of states at fixed level $n$ and angular momentum $J$, 
\begin{equation}
\label{eq:denstates}
    d_{n,J} = \frac{1}{2\pi i} \oint \frac{dx}{x^{n+1}} Z(x, J), \hspace{15pt} Z(x, J) = \int_{\mathbb{R}} \frac{dk}{2\pi} e^{-i k J} Z(x, k),
\end{equation}
where $k = i \beta \Omega$. 

To this effect, first we compute the partition function in the high-temperature limit $\beta\to 0$ at fixed angular velocity $\Omega$, using the modular transformation properties of the Jacobi theta and Dedekind eta functions, in terms of which the partition function \eqref{ZxOm} can be expressed 
\begin{equation}
\label{dosrot}
    Z(\beta \approx 0,\Omega) = \text{const.}\, \beta^{\frac{c}{2}} e^{\frac{c \pi^2}{6 \beta}} \frac{\Omega}{\sin(\pi \Omega)} =  Z(\beta \approx 0)\frac{\Omega}{\sin(\pi \Omega)} \,.
\end{equation}
The effect of adding angular momentum is thus an extra factor of $\Omega/\sin(\pi \Omega)$ compared to the static partition function.
The Fourier transform in \eqref{eq:denstates} can be performed exactly (see Appendix \ref{app:Fourier})
\begin{equation}
      Z(\beta\approx 0, J) = \text{const.}\,\beta^{b+1} e^{\frac{a}{\beta}}\sech^2\left(\frac{\beta |J|}{2}\right),
\end{equation}
where $a$ and $b$ were defined in \eqref{eq:defab}. We can then compute the density of states  
\begin{equation}
\label{dosrot5}
    d_{n,J} = \text{const.} \int d\beta \;e^{n\beta + \log{Z(\beta,J)}}, 
\end{equation}
using the saddle point approximation,
which leads to the condition
\begin{equation}
\label{saddlecond}
    n - \frac{a}{\beta^2} + \frac{b+1}{\beta} - |J| \tanh{\left(\frac{\beta |J|}{2}\right)} = 0.
\end{equation}
In the high-temperature limit $\beta\to 0$ we can solve \eqref{saddlecond} in the regimes where $\beta |J|$ tends to: zero, a finite constant value, and infinity, for which we can set
\begin{equation}
\label{sps}
    \tanh{\left(\frac{\beta |J|}{2}\right)} = \mu_d \in (0,1).
\end{equation}
The saddle point solution at large $n$ is then given by 
\begin{equation}
\label{betarotdos}
    \beta_* 
   {=}\sqrt{\frac{a}{n - \mu_d|J|}}
    +\order{n^{-1}}\,.
\end{equation}
The regimes $\beta|J|\to \{0,{\rm const},\infty\}$ are thus consistent with the scalings, at large excitation level $n$, of the angular momentum  $J=\{\order{1},\order{\sqrt{n}}, \order{n}\}$ which correspond to\footnote{Note that the regime $\beta|J|\to 0$, which yields saddle point $\beta_*=\sqrt{\frac{a}{n}}$, only captures the leading large $n$ behavior of the slowly rotating string which is independent of $J$ and thus the same as the static case; to capture the $|J|$-dependence of the density of states we will carry out a more refined analysis below.}
\begin{equation}
\label{mud}
    \mu_d =  \begin{cases}
       0, &  J = \order{1} \\
      \in (0,1),
      &  J = \order{\sqrt{n}} \\
      1, & J = \order{n}
    \end{cases}.
\end{equation}
The density of states at large excitation level $n$ then evaluates to
\begin{equation}\label{dos}
    d_{n, J} = \text{const.}\,(n - \mu_d|J|)^{-(D+3)/4} e^{\frac{\sqrt{a} (2 n - \mu_d|J|)}{\sqrt{n-\mu_d|J|}}}  \sech^2\Big[\frac{\sqrt{a} |J|}{2 \sqrt{n-\mu_d|J|}}\Big]. 
\end{equation}
This generalizes the earlier result of \cite{Russo:1994ev} for large $J= \order{n}$ to other scaling regimes.

\subsubsection{Size and shape}
\label{ssec:RotatingStringSize}

The main result of this article is the computation of the sizes of rotating strings parallel and transverse to the plane of rotation. 
The starting point is the same as in the static case \eqref{R2static}, that is, we compute 
\begin{equation}
    \langle R_{\perp}^2\rangle\equiv\frac{d}{dy}\Tr(x^N y^{R_{\perp}^2})\big|_{y=1} 
\end{equation}
from
\begin{equation}
\Tr(x^N y^{R_\perp^2})=\prod_{n=1}^\infty \left(\frac{1}{1-x^n}\right)^{D-5} \frac{1}{1-x^n y^\frac{1}{n}} \frac{1}{1-x^{n+\Omega}}\frac{1}{1-x^{n-\Omega}}\,.
\end{equation}
This yields a simple expression for the size transverse to the rotation plane
\begin{equation}
\label{perpsize}
   \langle R_\perp^2\rangle = \frac{1}{c} Z(\beta, \Omega) \log Z(\beta)\,.
\end{equation}
In the high-temperature limit this differs from  the static string result \eqref{statsize} simply by the multiplicative factor $\Omega/\sin(\pi \Omega)$ due to rotation. %
For the size parallel to the rotation plane,
\begin{equation}
    \langle R_{\parallel}^2\rangle\equiv\frac{d}{dy}\Tr(x^N y^{R_{\parallel}^2})\big|_{y=1} \,,
\end{equation}
we use
\begin{equation}
\Tr(x^N y^{R_\parallel^2})=\prod_{n=1}^\infty  \left(\frac{1}{1-x^n}\right)^{D-4}  \frac{1}{1-x^{n+\Omega}y^{\frac{1}{n}}} \frac{1}{1-x^{n-\Omega}y^{\frac{1}{n}}}\,, 
\end{equation}
and find that the result can be expressed as
\begin{equation}\label{parallsize}
    \langle R_\parallel^2 \rangle = \frac{1}{\beta} Z(\beta, \Omega) \rho(\beta,\Omega) \,,
\end{equation}
where
\begin{equation}\label{rhotation}
     \rho(\beta,\Omega)\equiv\beta \left(\frac{e^{-\beta(n+\Omega)}}{n(1 - e^{-\beta(n+\Omega)})} + \frac{e^{-\beta(n-\Omega)}}{n(1 - e^{-\beta(n-\Omega)})}\right).
\end{equation}
The leading contribution to \eqref{rhotation} in the high-temperature limit $\beta\to 0$ can be derived using a regularization procedure, which we discuss in Appendix~\ref{app:rhotation}. The result is the {\it rhotation function}
\begin{equation}
\label{rhohightemp}
       \rho(\beta\approx 0,\Omega)\approx \frac{1 - \pi \Omega \cot{(\pi \Omega)}}{ \Omega^2} .
\end{equation}

Finally, to compute the average sizes $\langle \bar r_\perp^2\rangle_n$ and $\langle \bar r_\parallel^2\rangle_n$ of the rotating string via \eqref{rsizes}, we perform yet another saddle point analysis. In the following, we only discuss the main results and refer to Appendix~\ref{app:Fourier} and Appendix~\ref{app:Sizes} for more details.

\paragraph{Size transverse to the rotation plane.}
The average size transverse to the plane of rotation,
\begin{align}
\label{eq:OrthoSize1}
    \langle \bar{r}^2_\perp \rangle_n \propto \frac{R^2_{n, \perp}}{d_{n, J}}, \qquad R^2_{n, \perp} = \frac{1}{2\pi i} \oint \frac{dx}{x^{n+1}} \langle R^2_\perp \rangle,
\end{align}
is computed by extracting the coefficient $R^2_{n, \perp}$ from \eqref{perpsize}. Up to the (subleading) factor $\frac{1}{c}\log Z(x)$, this calculation is essentially the same as the one for the density of states of the rotating string. The result is
\begin{equation}
    c\, R^2_{n, \perp} = \text{const.} \int d\beta e^{n\beta + \log{Z(\beta, J)} + \log\log Z(\beta)} = \log Z(\beta_*) \;d_{n, J},
\end{equation}
where the saddle point $\beta_*$ is given by \eqref{sps}.

For the average size of the string perpendicular to the rotation plane this yields
\begin{equation}\label{rperp}
     \frac{\langle \bar{r}^2_\perp \rangle_n}{\ell_s^2} \propto \sqrt{n -\mu_d |J| }=\begin{cases}
     \sqrt{n}, &  J = \order{1} \\
    \sqrt{n-\mu_d|J|} , &  J = \order{\sqrt{n}} \\
    \sqrt{n-|J|} , & J = \order{n}
    \end{cases},
\end{equation}
where we have reinstated string units, and the factor $\mu_d\in(0,1)$ is as in \eqref{mud}.
%
%
We see that the effect of rotation is to decrease the size of the string orthogonal to the plane of rotation to $\sim \sqrt{n-\mu_d |J|}$ compared to the static case where the size  $\sim \sqrt{n}$.

\paragraph{Size along the rotation plane.}
The computation of the average size in directions along the plane of rotation,
\begin{equation}
\label{rparaRpara}
    \langle \bar{r}^2_{\parallel}\rangle_n \propto \frac{R^2_{n, \parallel}}{d_{n, J}}\,, \qquad R^2_{n, \parallel} = \frac{1}{2\pi i } \oint \frac{dx}{x^{n+1}} \langle R^2_\parallel \rangle\,,
\end{equation}
with $ R^2_{n, \parallel}$ extracted from \eqref{parallsize}, requires more work. 
In the high-temperature limit $\beta\to 0$ at fixed angular velocity $\Omega$ we have
\begin{equation}
    \langle R^2_{\parallel} \rangle(\beta\approx 0,\Omega)=\text{const.} \,\beta^{b+1}e^{\frac{a}{\beta}} \frac{\Omega}{\sin(\pi \Omega)} \frac{1 - \pi \Omega \cot{(\pi \Omega)}}{\Omega^2} .
    \label{eq:parallelsize}
\end{equation}
where we use again $a$ and $b$ defined in \eqref{eq:defab}. The Fourier transform from angular velocity $\Omega$ to angular momentum $J$ can again be performed exactly by deforming the contour into the complex plane and evaluating the residues.
The result is
\begin{equation}
 \langle R^2_{\parallel} \rangle(\beta\approx 0,J)   =\text{const.} \,\beta^b e^{\frac{a}{\beta}} \left[\log{(1 + e^{-\beta|J|})} + \frac{\beta |J|}{1 + e^{\beta|J|}}\right]\,.
\end{equation}
We then compute
\begin{equation}
    R^2_{n, \parallel} = \text{const.} \int d\beta e^{n \beta + \log \langle R^2_{\parallel} \rangle(\beta,J)} 
\end{equation}
using the saddle point approximation, which yields the condition 
\begin{equation}
    n  - \frac{a}{\beta^2} + \frac{b}{\beta} - |J|\frac{\frac{\beta |J|}{4} \;\text{sech}^2{\left(\frac{\beta|J|}{2}\right)}}{\log{(1 + e^{-\beta|J|}) + \frac{\beta|J|}{1 + e^{\beta|J|}}}} = 0.
\end{equation}
We can again solve this equation for the regimes where $\beta |J|$ tends to zero, a finite constant value, or infinity, for which we can set
\begin{equation}
\label{eq:NuDef}
 \frac{\frac{\beta |J|}{4} \;\text{sech}^2{\left(\frac{\beta|J|}{2}\right)}}{\log{(1 + e^{-\beta|J|}) + \frac{\beta|J|}{1 + e^{\beta|J|}}}} = \mu_\parallel \in (0,1).
\end{equation} 
This is the same behavior as encountered in \eqref{sps} and so the saddle point is essentially given by \eqref{betarotdos}, with the replacement $\mu\to\nu$ and $b\to b-1$. Thus, the saddle point solution is 
\begin{equation}
  \label{betarotsize}
    \beta_{*} 
    =\sqrt{\frac{a}{n - \mu_\parallel|J|}} +\order{n^{-1}}\,.
\end{equation}
Again, the regimes of $\beta |J|\to \{0,{\rm const},\infty\}$ are consistent, at large excitation level $n$, with the three cases of interest \mbox{$J= \order{1},\order{\sqrt{n}},\order{n}$}. 
In these regimes, we approximate the prefactor of $|J|$ by
\begin{equation}
\mu_\parallel=  \begin{cases}
       0, &  J = \order{1} \\
      \in (0,1), 
      &  J = \order{\sqrt{n}} \\
      1, & J = \order{n}
    \end{cases}.
\end{equation} 
At large excitation level $n$ we find
\begin{equation}
    R_{n,\parallel}^2=\text{const.}(n - \mu_\parallel|J|)^{-(D+1)/4} e^{\frac{\sqrt{a} (2 n - \mu_\parallel|J|)}{\sqrt{n-\mu_\parallel|J|}}} \left[\log(1+e^{-\frac{\sqrt{a}|J|}{n-\mu_\parallel|J|}})+\frac{\frac{\sqrt{a}|J|}{n-\mu_\parallel|J|}}{1+e^{\frac{\sqrt{a}|J|}{\sqrt{n-\mu_\parallel|J|}}}} \right] .
\end{equation}
This finally yields, reinstating string units, the average size in directions along the rotation plane
\begin{equation}
\label{rparallel}
     \frac{\langle \bar{r}^2_\parallel \rangle_n}{\ell_s^2} \propto \begin{cases}
     \sqrt{n} 
     , &  J = \order{1} \\
    C_\parallel\sqrt{n-\mu_d|J|}
    , &  J = \order{\sqrt{n}} \\
     |J| , & J = \order{n}
    \end{cases},
\end{equation}
    where $\mu_d$ is again given by \eqref{mud} and
 \begin{equation}
 \label{Ceq}
    C_\parallel=\frac{1}{4\log 2}\left(1+e^{-\sqrt{\frac{a}{n}}|J|}\right)\left[\sqrt{\frac{a}{n}}|J|+\left(1+e^{\sqrt{\frac{a}{n}}|J|}\right)\log\left(1+e^{-\sqrt{\frac{a}{n}}|J|})\right)\right].
\end{equation}

We conclude that large angular momentum has the effect of dramatically increasing the size of the string in the directions along the rotation plane: from $\propto \sqrt{n}$ for the static string to $\propto |J|\sim n$ for the highly spinning string. 
For moderate spins there are two competing effects: The average size in the rotation plane decreases by a factor $\sqrt{n-\mu_d |J|}$ from dividing by the density of states $d_{n,J}$ in \eqref{rparaRpara}, while the size coefficient $R^2_{n,\parallel}$ scales up the average size in the rotation plane by a factor of $C_\parallel>1$. Physically, the two competing effects are the centrifugal expansion due to the rotation, and the reduction in random-walk spread due to fewer random oscillators.

\paragraph{Sizes beyond the leading order in large excitation level $n$.}
To capture the dependence of the size on angular momentum for small $J=\order{1}$ we approximate the saddle point equations by
\begin{equation}
n-\frac{a}{\beta^2}+\frac{B}{\beta}-\#\beta J^2=0,
\end{equation}
where $\#=\frac{1}{2}$ for the saddle point equation that computes the density of states $d_{n,J}$  where $B=b+1$, while $\#=\frac{1}{2}$ and $\#=1/(4\log 2)$, respectively, for the ones that compute the sizes transverse to and along the rotation plane, $R^2_{n,\perp}$ and $R^2_{n,\parallel}$  where $B=b$.

Repeating the above analyses,
we find that the average sizes transverse to and along the rotation plane are 
\begin{equation}
\label{rperpsmallJ}
     \frac{\langle \bar{r}^2_\perp \rangle_n}{\ell_s^2} \propto \sqrt{\frac{n}{a}}+\gamma_1+\frac{\gamma_2}{\sqrt{n}}+\order{n^{-1}},
\end{equation}
and
\begin{equation}
\label{rparallelsmallJ}
      \frac{\langle \bar{r}^2_\parallel \rangle_n}{\ell_s^2} \propto \sqrt{\frac{n}{a}}+\gamma_1+\frac{\gamma_2}{\sqrt{n}}+J^2 \frac{\gamma_\parallel}{\sqrt{n}}+\order{n^{-1}}.
\end{equation}
The constants $\gamma_1$, $\gamma_2$ and $\gamma_\parallel$ are given by
\begin{equation}
\label{eq:consts}
    \gamma_1=\frac{3+2b}{4a},\qquad \gamma_2=\frac{11+4b(4+b)}{32a\sqrt{a}}, \qquad \gamma_\parallel=\sqrt{a}\frac{2\log 2-1}{8\log 2}.
\end{equation}
For slowly spinning strings with $J=\order{1}$ the effect of rotation kicks in at order $\order{n^{-1/2}}$, where it increases the size of the string in the directions along the plane of rotation.
Notice that the expressions \eqref{rperpsmallJ} and \eqref{rparallelsmallJ} agree for the first few orders in the large $n$ expansion, and only start to differ at order $\order{n^{-1/2}}$. This leads to a delicate cancellation when computing the ratio of sizes transverse to and along the plane of rotation. We turn to it next.

\paragraph{Size ratios.}

Self-gravitation will significantly affect the sizes of the string, but this effect should be roughly the same in all directions, at least in the regime of small and moderate spins, where the string is expected to be fairly isotropic on average.
Therefore, we can reasonably expect that the ratio of sizes transverse to and along the plane of rotation is an approximate adiabatic invariant that can be meaningfully compared for objects on different sides of the correspondence point.
For the ratio of string sizes, we find the following behavior:\footnote{In our results below we have ignored the overall $\order{1}$ constant prefactors of ${\langle \bar{r}^2_\perp \rangle_n}/{\ell_s^2}$ and $ {\langle \bar{r}^2_\parallel \rangle_n}/{\ell_s^2}$. One would expect that these factors cancel in the size ratio in the static limit $J\to 0$. In Appendix~\ref{subsec:overratio} we show that they do not and we explain this fact and why it does not affect our conclusion.}

\begin{itemize}
    \item {\bf $J = \order{1}$:} For strings with small angular momentum, the ratio of sizes orthogonal to and along the plane of rotation is
   \begin{equation}
   \label{eq:SmallJRatio}
    \frac{\langle \bar{r}^2_\perp \rangle_n}{\langle \bar{r}^2_\parallel \rangle_n} \propto 
    1-\gamma_\parallel \frac{J^2}{S^2},
\end{equation}
where we note that $\gamma_\parallel>1$.
    We observe that the effect of rotation for a slowly rotating string is to increase its size in directions along the plane of rotation leading to a smaller size ratio compared to the static case. 
    Moreover, 
    the first correction in the spin is of the form $\frac{J^2}{S^2}$. As we explained in the introduction, this is significant,  and it matches the behavior of a slowly rotating black hole. 

    \item {\bf $J = \order{\sqrt{n}}$:} For strings with moderate angular momentum we find 
       \begin{equation}\label{modratio}
    \frac{\langle \bar{r}^2_\perp \rangle_n}{\langle \bar{r}^2_\parallel \rangle_n} \propto \frac{1}{C_\parallel}<1.
\end{equation}
The effect of moderate rotation on the string size is relatively modest: it scales up the size of the string along the rotation plane relative to the one transverse to it by a factor $C_\parallel>1$ that varies slowly with $J/\sqrt{n}=\order{1}$.
    \item {\bf $J = \order{n}$:} For strings with large angular momentum we find
    \begin{equation}
    \frac{\langle \bar{r}^2_\perp \rangle_n}{\langle \bar{r}^2_\parallel \rangle_n} \propto 
     \frac{\sqrt{n-|J|}}{|J|}\propto   \frac{S}{|J|}.
    \end{equation}
    We observe a {\it pancake effect}: compared to a static string a highly spinning string has a larger size in direction along the rotation plane and a smaller size transverse to it. We will see below that this effect is much larger for rotating strings than for black holes. However, as discussed in \cite{Ceplak:2023afb}, we do not expect highly spinning strings and black holes to be in direct correspondence. The discrepancy is simply a consequence of the very different shapes of the two objects at large angular momenta.
\end{itemize}

Although we presented the results for the size of a rotating open string, one can show that the same qualitative statements hold for the closed string as well. Detailed calculations can be found in Appendix~\ref{sec:detsrot}, from which we can see that the final results are the same as for the open string case, with a change $|J| \to |J_L| + |J_R|$. We can write $J_i = c_i J$, where $c_i$ determine the value of the angular momenta,  since all of our results assume that $J_L$ and $J_R$ scale in the same way with $n$, and then the results explicitly take the form outlined above.


\section{Size and shape of rotating black holes}
\label{sec:BHSize}

Having computed the size and shape of rotating strings as a function of their level $n$ and spin $J$, let us now obtain them for rotating black holes as a function of their entropy~$S$ and angular momentum~$J$. Gravity introduces a stronger dependence on the number $D$ of spacetime dimensions, since four-dimensional black holes only exist when $2\pi|J|\leq S$ while in $D\geq 5$ black holes can have $|J|\gg S$. Throughout our discussion, we only consider black holes with rotation on a single plane.

\subsection{Basic properties}

We write the metric of the Myers-Perry black hole \cite{Myers:1986un} in the form
\begin{align}\label{MPbh}
    ds^2=&-dt^2+\frac{r_0^2+a_J^2}{\Sigma}\left(\frac{r_0}{r}\right)^{D-5}\left(dt+a_J\sin^2\theta\, d\phi\right)^2+\Sigma\left(\frac{dr^2}{\Delta}+d\theta^2\right)+(r^2+a_J^2)\sin^2\theta d\phi^2\nonumber\\
    &+r^2\cos^2\theta\, d\Omega^2_{D-4}\,,
\end{align}
with
\begin{align}
    \Sigma=r^2+a_J^2\cos^2\theta\,,\qquad \Delta =r^2+a_J^2-\left(\frac{r_0}{r}\right)^{D-5}(r_0^2+a_J^2)\,.
\end{align}
To translate the horizon radius $r_0$ and spin length $a_J$ into entropy $S$ and angular momentum $J$ we use
\begin{align}\label{r0aSJ}
   r_0=\frac{D-2}{4\pi}\frac{S}{M}\,,\qquad  a_J=\frac{D-2}{2}\frac{J}{M}\,,
\end{align}
with the black hole mass being
\begin{align}\label{MSJ}
    M=c_D\, S^{\frac{D-5}{D-2}}\left(S^2+4\pi^2 J^2\right)^{\frac1{D-2}}\,,\qquad c_D=\frac{D-2}{4\pi}\left(\frac{\Omega_{D-2}}{4}\right)^{\frac1{D-2}}\,.
\end{align}
We have set $G=1$ and $\hbar=1$ so that all dimensionful quantities are measured in Planck units. The simple, $D$-independent relation
\begin{align}
    \frac{r_0}{a_J}=\frac{S}{2\pi J}
\end{align}
is useful and implies that the ultraspinning regime $J\gg S$ is reached by taking $a_J\gg r_0$. Observe also that, for a given entropy $S$, the mass $M$ always grows with $J$ since kinetic energy is added to the `heat' energy of the black hole (also known as irreducible mass).

\subsection{Sizes} 

The size and shape of the black hole are more difficult and ambiguous to characterize than those of the string because the geometry is strongly distorted by curvature. Here we will follow and slightly extend the analysis in \cite{Emparan:2003sy}.

Unlike the case of the string, where we separate the size along the two-dimensional rotation plane, and along the $D-3$ transverse spatial directions, in the black hole metric \eqref{MPbh} it is more convenient to first separate the $D-4$ directions $\Omega_{D-4}$ that are orthogonal to the rotation plane and to the polar angle $\theta$. The area of the horizon in these directions,
\begin{align}
    \mathcal{A}^{(D-4)}_\perp=\Omega_{D-4} (r_0\cos\theta)^{D-4}
\end{align}
varies along the polar angle $\theta$. If we define $r_\perp$ as the area-radius at the pole $\theta=0$, then
\begin{align}
    r_\perp =r_0\,.
\end{align}

For the size in directions along the rotation plane, the choice is less obvious. One possibility is to measure the proper length of the horizon equator at $\theta=\pi/2$ and define
\begin{align}
    r_\textrm{eq} \equiv \sqrt{g_{\phi\phi}}|_{r=r_0,\theta=\frac{\pi}2}=r_0+\frac{a_J^2}{r_0}\,.
\end{align}
However, in the ultraspinning regime, this behaves like $ r_\textrm{eq}\simeq a_J^2/r_0$, which is much larger than the characteristic parallel size $\sim a_J$ used in \cite{Emparan:2003sy}. The latter behavior can be motivated in several ways. One can compute the area-radius of the horizon  in the $(\theta,\phi)$ directions at fixed $\Omega_{D-4}$,
so that
\begin{align}
    r_\|^{(a)} \equiv \sqrt{\frac{\mathcal{A}^{(2)}_\|}{4\pi}}=\sqrt{r_0^2+a_J^2}\,.
\end{align}
A more physically-motivated choice is the critical impact parameter $r_\|^{(b)}$ for the capture of a null geodesic in the equatorial plane, which in \cite{Emparan:2003sy} was found to be given by
\begin{align}
    \left(r_\|^{(b)}+a_J\right)^{D-1} \left(r_\|^{(b)}-a_J\right)^{D-5}=\frac14\frac{(D-1)^{D-1}}{(D-3)^{D-3}}r_0^{2(D-3)}\left(1+\frac{a_J^2}{r_0^2}\right)^2\,.
\end{align}
Up to a numerical factor, both $ r_\|^{(a)}$ and $r_\|^{(b)}$ scale like $a_J$ in the ultraspinning regime and behave very similarly more generally. Since $ r_\|^{(a)}$ has a much simpler expression, in the following we will use it for $ r_\|$ instead of $r_\|^{(b)}$, and also discard $r_\textrm{eq}$. Let us note, however, that in the regime of $|J|\ll S$ relevant to the correspondence, all these choices give equivalent results up to purely numerical factors.

Now we write these size radii in the form
\begin{align}
    r_\perp=&\frac{D-2}{4\pi}\frac{S}{M}\,,\label{RperpSM}\\
    r_\|=& \frac{D-2}{4\pi}\frac{\sqrt{S^2+4\pi^2 J^2}}{M}\label{RparaSM}\,,
\end{align}
with $M(S,J)$ given in \eqref{MSJ}. Then the ratio we are interested in takes a simple exact form
\begin{align}
    \frac{r_\perp^2}{r_\|^2}=\frac{S^2}{S^2+4\pi^2 J^2}\,.
\end{align}
From here, the results in \eqref{eq:bhratios} for the different spinning regimes immediately follow. The fact that this ratio is independent of $D$ is not very significant; the alternative choices discussed above are slightly more complex and, in the small spin regime, yield the same results but with $D$-dependent factors.

Black rings with horizon topology $S^1\times S^{D-3}$ exist in $D\geq 5$ but their spin is bounded below and cannot reach $|J|\ll S$. Their properties to leading order for large $|J|\gg S$ are known in all $D\geq 5$ \cite{Emparan:2007wm}. Associating the horizon radii of the $S^1$ and $S^{D-3}$ factors
with $r_\|$ and $r_\perp$, respectively, one finds
\begin{align}
    r_\perp^{D-2}\sim\, \frac{S^2}{|J|}\,,\qquad
    r_\|^{D-2}\sim \, \frac{|J|^{D-3}}{S^{D-4}}\,.
\end{align}
Compared with a Myers-Perry black hole with the same entropy, a black ring is fatter and longer, as could be expected on simple grounds. However, we concluded in \cite{Ceplak:2023afb} that neutral black rings do not have a correspondence with fundamental string states, so these size ratios cannot be meaningfully compared with our results for rotating strings.

Rotating black bars can play a role in the correspondence in the regime of $|J|\sim S$, but since they are only long-lived and not stationary configurations, our information about their properties is very limited. We know that in the large-$D$ limit, and for $|J|\gg S$, the results for their sizes are parametrically the same as for Myers-Perry black holes, i.e., $r_\perp \sim \order{1}$ and $r_\|\sim |J|/S$ \cite{Andrade:2018nsz}. However, this is an ultraspinning regime that is outside the correspondence with fundamental string states.

\section{Random walk model for rotating strings}
\label{sec:randomwalk}

The picture of highly excited strings as random walks is especially useful when interpreting their more geometric properties, such as their average size\cite{Mitchell:1987hr, Mitchell:1987th}.
In this section we construct a random walk model which completely reproduces the sizes of strings at fixed level $n$ and  angular momentum $J$, as derived in Section~\ref{sec:sizes}.
When applied to the static case, the random walk model reproduces the results of \cite{Mitchell:1987hr, Mitchell:1987th, Manes:2004nd}.
Using this model, we can interpret the scaling for the sizes of rotating strings in the plane of rotation and transverse to it.
In the following, we focus on the qualitative setup of the model, while leaving the detailed calculations to Appendix~\ref{app:RandomWalk}.

\subsection{Random walk as a path integral}
\label{ssec:RWSetup}

The goal of this section is to model highly excited closed strings as (a pair of) closed random walks in the $c=D-2$ dimensions orthogonal to the worldsheet, where $D$ is the number of spacetime dimensions.
The most general closed string solution is a sum of left and right movers
\begin{align}
\label{eq:ClosedStringSum}
    X_i(\tau, \sigma) = X_i^L(\tau +\sigma) + X_i^R(\tau-\sigma)\,,
\end{align}
with $(\tau, \sigma)$ being the world-sheet coordinates. 
The oscillator modes, $\alpha_n$ and $\tilde{\alpha}_n$, associated to each sector (see \eqref{Xclosed0}) are independent, up to the level matching condition.
When written in terms of these oscillator modes, the total angular momentum  is just a sum of the left and right contributions, which we denote $J_L$ and $J_R$ respectively. 
Indeed, we can write the closed string analogue of the Hamiltonian \eqref{eq:Hamil1} as 
\begin{align}
    H^{\rm closed}  = N_L + N_R - \Omega_L\,J_L - \Omega_R\,J_R\,,
\end{align}
where we have introduced separate angular potentials $\Omega_{L,R}$ for each sector and  
\begin{align}
\label{eq:NJOpDef}
    N_L = \sum_{n=1}^{\infty}\sum_{i=1}^{c} \,\alpha_{-n}^i\,\alpha_{n}^i\,,\qquad J_L = -i\sum_{n=1}^\infty \frac{1}{n} \left(\alpha_{-n}^1 \alpha_n^2-\alpha_{-n}^2 \alpha_n^1\right)\,,
\end{align}
with their right-moving counterparts defined in an analogous way using $\tilde \alpha_n$.
Similarly, we can write the closed string partition function as  the product of two open string partition functions,
\begin{align}
\label{eq:CLO2}
    Z_{\rm closed}(\beta, k_L,k_R) 
     = Z_{\rm open}(\beta, k_L)\times Z_{\rm open}(\beta, k_R)\,,
\end{align}
with $k_{L,R} = i\,\beta\,\Omega_{L,R}$, subject to the level matching condition $N_L = N_R$.

The underlying geometric idea that we introduce for the random walk model is that highly excited closed strings are described by two collections of independent random walks, one for the left-movers and another for the right-movers.%
\footnote{Our setup is inspired by the random walk model in \cite{Manes:2004nd}, but with some important differences discussed in Appendix~\ref{sapp:CanEns}. Similar ideas have been explored in the context of BPS configurations in \cite{Martinec:2023xvf, Martinec:2024emf}.} The total length, $L$, is proportional to the mass, $M$, in string units
\begin{align}
    L \propto \frac{M}{M_s}\,\ell_s \propto \sqrt{n}\,\ell_s\,.
\end{align}
We have used that the mass scales as the square root of the excitation number $n$ for highly excited strings. 
For simplicity, here we focus only on a single random walk which we compare to the sizes calculated in Section~\ref{sec:sizes}. 
The generalisation to a pair of random walks describing the full closed string solution is presented in Appendix~\ref{sapp:Closed}.

Random walks sweep out a density profile in $\mathbb{R}^c$.
As in \cite{Manes:2004nd}, where the authors also use a random walk to find a density distribution of closed strings, we parametrize closed random walks that go through a chosen point %
$x_i\in \mathbb{R}^c$ as smooth curves $X_i(s)$ in $\mathbb{R}^c$, where $0 \leq s\leq \pi$,%
\footnote{In Appendix~\ref{app:RandomWalk} we consider a general interval $0\leq s\leq \ell$, but show that to match the results obtained by the operator method, one needs to set $\ell =\pi$.  This is consistent with the prefactor in front of the first term in the Hamiltonian \eqref{eq:Hamil1}.}
that start and end at the point of interest, $x_i$,
\begin{align}
\label{eq:BC}
    X_i(0)= X_i(\pi) = x_i\,.
\end{align}
The total Euclidean length of the random walk is given by 
\begin{align}
    L[X_i] =  \int_0^\pi\,\sqrt{\dot{X}_i\,\dot{X}^i}\,ds\,,
\end{align}
where $\dot{}=d/ds$. 
However, since $L\propto \sqrt{n}$ and the above expression contains an unruly square root, it is more convenient to study the square of the length elements, which we naturally denote as
\begin{align}
\label{eq:Ndef}
    n\left[X_i\right]= \frac{1}{4\pi\,\alpha'}\,\int_0^\pi\,\dot{X}_i\,\dot{X}^i\,ds\,.
\end{align}
This gives the familiar kinetic term in non-relativistic classical mechanics. 
The normalization factor is chosen to emphasize the reminiscence to the 
Polyakov world-sheet action.

We are interested in string configuration with a single non-vanishing angular momentum $J$ in the 1-2 plane.
In the random walk model, we need a quantity that measures the analogue of the angular momentum of an oscillating string.
The natural candidate is 
\begin{align}
\label{eq:Jdef}
    J\left[X_i\right] = \frac{1}{2\pi\,\alpha'}\,\int_0^\pi\left(X_1\,\dot X_2 - X_2\,\dot X_1\right)\,ds\,.
\end{align}
Namely, the flat space and the (square of the) length are invariant under the rotation group $SO(c)$. 
It is then natural to treat \eqref{eq:Ndef} as a Euclidean action and construct conserved charges using Noether's theorem. 
Indeed, $J[X_i]$ defined in the above way is the conserved charge related to the rotation symmetry in the 1-2 plane.

\begin{figure}[t]
    \centering
    \includegraphics[width=0.9\linewidth]{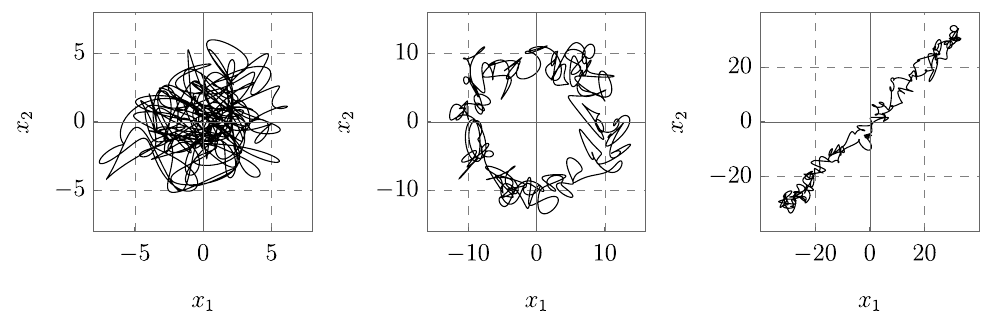}
    \caption{\small Examples of snapshots of string configurations projected onto the 1-2 plane. In the leftmost panel, we have a string ball, which on average moves as many times in one direction as in the other. Correspondingly, the area of such ball-like configurations is small. For the plot shown here $\alpha'\,J[X_i]\approx 4.13$.
    In the middle and rightmost panel, we have two configurations where the random walk moves predominantly in one direction, with small deviations. Such ring and bar-like configurations have a larger area, and indeed, for the configurations pictured, $\alpha'\,J[X_i]\approx  80.3$ for the ring and $\alpha'\,J[X_i]\approx  52.2$ for the bar. 
    We expect such configurations to give the leading contribution to the path integral at large values of the area. 
    The plots are generated using \cite{Karliner:1988hd}. }
    \label{fig:JStrings}
\end{figure}

This definition also provides a heuristic geometric interpretation: If $(X_1(s), X_2(s))$ were to describe a simply connected curve in two-dimensions, then \eqref{eq:Jdef} would be proportional to the area enclosed by the curve. 
Naturally, the projection of a generic random walk onto the 1-2 plane does not lead to a simply connected curve, but we can still adopt this intuition: By labelling different random walks according to their values of $J[X_i]$ we roughly separate them based on the area enclosed by the $(X_1(s), X_2(s))$ curve in the 1-2 plane, see Figure~\ref{fig:JStrings}.%
\footnote{Similar ideas have already been explored in two-dimensional Polymer physics, see for example \cite{Khandekar_1988, Duplantier_1989, Comtet_1990}. We thank Jean-Marc Luck for bringing our attention to these references.}
With that in mind, we refer to \eqref{eq:Jdef} as the area of the string in the 1-2 plane. Observe also that, just like a non-zero angular momentum forces a string to spread along the rotation plane, fixing the area of the walk on a plane also has the effect of imposing a minimal spread of the walk on that plane.

Let us denote the number
of random walks that pass through a point $x$  with square length equal to $n$ and area equal to $J$  by $d_{n,J}(x_i)$.\footnote{The number of smooth random walks passing through a point is infinite. To render it finite, a discretization or quantization must be introduced. We will elaborate on this shortly.} 
We can express this quantity as a sum over all possible closed paths that go through point $x_i$ as
\begin{align}
\label{eq:Density2}
d_{n,J}(x_i) \propto \sum_{P_c(x_i)}\delta\left(n 
- n\left[X_i\right]\right) \,\delta\left(J-J\left[X_i\right]\right)\,\delta\left(\avg{X_i}\right)\,,
\end{align}
where we have also imposed that the center of mass of the path
\begin{align}
\label{eq:AvgXdef}
    \left\langle X_i\right\rangle = \int_0^\pi X_i\,ds\,,
\end{align}
is at the origin, since we are interested only in the spread of the random walk away from its center of mass.
Rewriting the Dirac deltas as integrals of exponential functions and using  \eqref{eq:Ndef} and \eqref{eq:Jdef}, we can express the sum over all closed paths as a path integral 
\begin{align}
    \label{eq:PIRot}
    d_{n,J}(x_i) =   \int_{X_i(0) = x_i}^{X_i(\pi) =x_i}\,\cD X_i\,\int\,d\beta \,e^{\beta\,n}\,e^{\frac{a}{\beta}}\,\int\,dk\,e^{-i\,k\,J}\int\,d\mu_i\,e^{-I_k[X_i]}\,,
\end{align}
where we sum over all paths starting and ending at $x_i$ weighted by the action
\begin{align}
\label{eq:Action1}
     I_k[X_i] = \frac{1}{{4\pi\,\alpha'}}\int_0^{\pi}\,\left[{\beta}\dot X_i\,\dot X^i+ 4\pi i\,\alpha'\,\mu_i\,X_i- 2i\, k\left(X_1\,\dot X_2 - X_2 \,\dot X_1\right)\right]\,ds\,.
\end{align}
Note that we have introduced an additional normalisation factor $e^{a/\beta}$, which we cannot derive just from considering the sum over closed random walks \eqref{eq:Density2} alone. This is important,
and we will comment on it at the end of this subsection. For now, let us mention that its appearance is related to the fact that we are approximating an inherently quantum object using smooth classical paths and this normalisation factor provides an effective quantisation for the random walks.

To evaluate \eqref{eq:PIRot}, we assume that one can exchange the order of integration and first evaluate the path integral with $\beta$, $\mu_i$, and $k$ acting as free parameters.%
\footnote{We believe this to be justified, since the final result after exchanging the order of integration is convergent and finite. However, one may need to be more careful, since one of the integrals involved is a path integral.}
We recognise in \eqref{eq:Action1} the action of a non-relativistic charged particle moving in an electric and magnetic field. 
A path integral with such an action can be evaluated exactly using standard techniques \cite{feynman2010quantum}, with the details shown in Appendix~\ref{app:RandomWalk}. 
After performing the integral over the Lagrange multipliers for the center of mass condition $d\mu_i$, we are left only with, up to unimportant constant prefactors,
\begin{equation}
\label{eq:RWDensity}
    d_{n,J}(x_i) = \int d\beta \,e^{\beta\,n}\int\,dk\,e^{-ikJ}e^{\frac{a}{\beta}}\,\beta^{\frac{c}{2}}\,\left(\frac{3\,\beta}{\pi^3\,\alpha'}\right)^{\frac{c}{2}}\!\!\frac{1}{3\rb\left(\frac{\pi\,k}{\beta}\right)}\,\frac{\frac{\pi\,k}{\beta}}{\sinh\left(\frac{\pi\,k}{\beta}\right)} \exp\left[-\frac{3\,\beta}{\pi^2\,\alpha'}\left(\frac{x_1^2 + x_2^2}{3\,\rb\left(\frac{\pi\,k}{\beta}\right)}+x_{\perp}^2\right)\right]\,,
\end{equation}
where $x_{\perp}^2 = x_3^2 +\ldots x_c^2$.
In this expression we have introduced the function
\begin{align}
    \rb(y) \equiv \frac{\coth y}{y}- \frac{1}{y^2}\,,
\end{align}
which is, up to an overall constant factor, the same as the rhotation function \eqref{rhohightemp} evaluated at imaginary values of the argument.
Note that the spatial distribution of random walks has a Gaussian profile, with the size in the orthogonal directions unaffected by the presence of the area condition.%
\footnote{A similar distribution was found in \cite{Martinec:2023xvf}, which used random walks to model  BPS configurations in Type II strings and found that when expressed in terms of angular velocity $\Omega$ (in our case $k = i\beta\,\Omega$), the spatial distribution is also Gaussian. For configurations with fixed $J$, the spatial distribution becomes ring-like, which is expected for the BPS states under investigation.
It would be interesting to analyse this in our setup, where the dominant stringy configurations at large $J$ are expected to look like bars, although rings should also exist \cite{Blanco-Pillado:2007eit}.}

In practice we are interested in expectation values
\begin{align}
\label{eq:ObseDef}
    \avg{A}_{n,J} \equiv \frac{\int\,dx_i\,A\,d_{n,J}(x_i)}{\int \,dx_i \,d_{n,J}(x_i)}\,,
\end{align}
which involve integrals of \eqref{eq:RWDensity} over the entire space. 
Assuming that one can first perform the $dx_i$ integrals, one finds that the remaining integral transforms can always be evaluated using saddle point methods. 

\paragraph{The importance of the normalisation factor.}
When constructing \eqref{eq:RWDensity}, the most important point at which one relates the random walk with the fundamental string is through the normalisation factor $e^{a/\beta}$.\footnote{There are also prefactors of the integrals in \eqref{eq:Ndef} and \eqref{eq:Jdef}, but these determine only the constant overall prefactors of the size, and not the relevant scaling with $n$ and $J$.} 
This factor is determined by imposing that \emph{the integral of $d_{n,J}(x_i)$ over the entire space is equal to the number of strings at level $n$ and angular momentum $J$ in the thermodynamic limit:}
\begin{align}
\label{eq:Normalisation}
    \int\,dx_i\,d_{n,J}(x_i) = \int d\beta \,e^{\beta\,n}\int\,dk\,e^{-ikJ}\,\beta^{\frac{c}{2}}\,e^{\frac{a}{\beta}}\,\frac{\frac{\pi\,k}{\beta}}{\sinh\left(\frac{\pi\,k}{\beta}\right)}  \equiv d_{n,J}\,,
\end{align}
where the expression for the right-hand side is given by \eqref{dosrot5}.
This expression for $d_{n,J}$ is an external input that needs to be provided for the random walk to be consistent with the string calculation.
As we show in Appendix~\ref{app:RandomWalk}, one actually only needs to fix the exponential factor $e^{a/\beta}$, because  $\beta^{c/2}$ and the $k$-dependent function appear naturally in the one-loop determinant of the path integral in \eqref{eq:PIRot}.

The identification \eqref{eq:Normalisation} seems reasonable, since $d_{n,J}(x_i)$ counts the number of random walks through a fixed point $x_i$, so the integral over the entire space should reproduce the number of all closed random walks, which we then relate to the number of highly excited string states.
Indeed, in the argument of the  Fourier and Laplace transforms we recognise the partition function of rotating open strings in the thermodynamic limit \eqref{dosrot}, with $k =i\,\beta\,\Omega$.
This allows us to interpret $\beta$, that was introduced in the random walk model, as the conjugate variable of the level number $n$ and the Lagrange multiplier, $k$, imposing the area $J[X_i]=J$, as  being related to the angular potential $\Omega$.
This identification also provides a limitation on the random walk model: We can only trust the random walk at large temperatures, where the partition function is given by \eqref{dosrot}. 

The fact that $e^{a/\beta}$ needs to be put in by hand is not surprising.
In our model, we sum over smooth classical paths and then want to impose that their number equals the number of highly excited quantum string states.  
By adding this normalisation factor, we effectively introduce a quantisation for the path integral. 
As can be seen in the saddle point calculations in \eqref{staticsaddle} or \eqref{saddlecond}, this factor is crucial to ensure the correct exponential growth of the number of states with $n$ and $J$.
In turn, this exponential growth is exactly the leading degeneracy of random walks with $\sqrt{n}$ steps of string length $\ell_s$ \cite{Horowitz:1997jc, Barbon:2004dd}. 
One could thus interpret the addition of this normalisation factor as a way to ensure that the exponential growth of the number of random walks is correct, without requiring the knowledge of the partition function of string states. 

A different perspective into this normalisation factor is through the lens of modular invariance.
The partition function of the string states is modular invariant and, in fact, its  leading high-temperature behaviour \eqref{dosrot} is obtained using modular transformations.
The random walk model, on its own, lacks any intrinsic information about modular invariance. However, introducing the normalization factor $e^{{a}/{\beta}}$ restores the correct leading-order behavior at high temperatures.
It would be very interesting to investigate whether imposing modular invariance on a random walk model would naturally incorporate this normalization factor, independent of any reference to the partition function of quantized strings. In this respect, see also \cite{Kruczenski:2005pj,Barbon:2004dd,Mertens:2015ola}.

Finally, we obtain $d_{n,J}(x_i)$ by ``counting'' the number of closed random walks that begin and end at $x_i$. This may lead to an overestimation or underestimation of the actual number of random walks. Introducing the normalization factor helps to mitigate these inaccuracies.

\paragraph{Closed string boundary condition and the open string partition function.}
In the equation \eqref{eq:Normalisation} we have matched the number of \emph{closed} random walks to the number of  \emph{open} strings.%
\footnote{Boundary conditions for closed paths are easier to implement than those of open paths in the random walk model.}
This is justified because we are always implicitly considering one of the two independent modes of closed strings \eqref{eq:ClosedStringSum}. 
As we have shown in \eqref{eq:CLO2}, the partition function of the closed string is the product of two open string partition functions at equal temperature.
Then, it is reasonable to compare the random walk model of an individual sector with the partition function of the open string, despite the clash between the boundary conditions.
In addition, we show in the next subsection that one random walk exactly reproduces the sizes of rotating open strings.
We show in Appendix~\ref{sapp:Closed} how an appropriate sum of two independent random walks exactly reproduces the closed string partition function and the corresponding size calculations for closed strings.

\paragraph{Summary and heuristic interpretation.} 
In this random walk model, we employ a Euclidean path integral to sum over all possible closed paths with a fixed squared length and fixed area, yielding a density distribution of strings at a given spatial point. This approach enables the calculation of moments of the spatial distribution, which, in turn, allows us to determine the expected size of rotating strings.

In our derivation, we have imposed that the integral of the density distribution of random walks over the entire space is equal to the number of strings at fixed level $n$ and angular momentum $J$.
This is equivalent to identifying%
\footnote{
We have discussed the subtleties between the different boundary conditions in the previous paragraph.}
\begin{align}
    Z(\beta,k)={\rm Tr}\left(e^{-\beta\,N - i \,k\,J}\right)\equiv  \int\,dx_i\,\int_{X_i(0) = x_i}^{X_i(\pi) =x_i}\,\cD X_i\,e^{\frac{a}{\beta}}\int\,d\mu_i\,e^{-I_k[X_i]}\,,
\end{align}
where the action is given by \eqref{eq:Action1} and the operators $N$ and $J$ are defined in \eqref{eq:NJOpDef}.
By comparing the two exponential functions, it is then tempting to propose a translation
\begin{subequations}
\label{eq:Translation}
    \begin{align}
    &N = \sum_{n=1}^{\infty}\sum_{i=1}^{c} \,\alpha_{-n}^i\,\alpha_{n}^i&&\Longleftrightarrow && n[X]=\frac{1}{4\pi\,\alpha'}\int_{0}^{\pi}\,\dot X_i\,\dot X^i\,ds\,,\\
    &J = -i\sum_{n=1}^\infty \frac{1}{n} (\alpha_{-n}^1 \alpha_n^2-\alpha_{-n}^2 \alpha_n^1) &&\Longleftrightarrow && J[X]= \frac{1}{2\pi\,\alpha'}\,\int_0^\pi\left(X_1\,\dot X_2 - X_2\,\dot X_1\right)\,ds\,.
\end{align}
\end{subequations}
This puts flesh on the statement that highly excited strings, generated by repeated action of the $\alpha_{-n}^i$ oscillator modes, can be modeled by a random walk. 

Finally, by integrating over all values of $J$, or equivalently setting $\Omega = 0 $, one finds a random walk model for static strings.
Integrating such a model over all space exactly reproduces the density of states of static strings and, following the same procedure as outlined in the next subsection, one finds that the size of the random walk exactly matches the expected size of static strings  \eqref{eq:StaticSize} \cite{Mitchell:1987hr, Mitchell:1987th, Manes:2004nd}.

\subsection{Sizes of random walks at fixed \texorpdfstring{$n$}{n} and \texorpdfstring{$J$}{J}}
\label{ssec:RWSizes}

We now calculate the expectation value of the size of the random walk model in the directions orthogonal to the plane 1-2 and in this plane. 
We show that these sizes exactly match the sizes of rotating strings in the thermodynamic limit presented in Section~\ref{sec:sizes}. 
In particular, we show that the random walk model exactly reproduces the integrals of \eqref{perpsize} and \eqref{eq:parallelsize}, meaning that the sizes obtained through the random walk model agree for all regimes of the angular momentum $J$.

These results provide compelling evidence that highly excited strings exhibit random walk behavior, demonstrating that the phenomenon holds true for arbitrary values of the angular momentum.

\paragraph{Size transverse to the rotation plane.}

Using the definition of expectation values \eqref{eq:ObseDef} in the random walk model, we can look at the expected value of an arbitrary direction orthogonal to the plane of rotation, $x_{\perp}$.
Since we have set the center of mass to be at the origin, the first moments vanish, $\avg{x_i}=0$.  
The non-trivial moment of the Gaussian distribution is
\begin{align}
\label{eq:DefOrthoSize}
    \avg{x_{\perp}^2}_{n,J} = \frac{1}{d_{n,J}}\,\int\,dx\,x_{\perp}^2\,d_{n,J}(x_i)\,,
\end{align}
where $d_{n,J}$ is given in \eqref{eq:Normalisation} and is explicitly evaluated in \eqref{dos}.
To evaluate the integral in the numerator, we assume that we can exchange the order of integration and perform the Gaussian integral first, after which we get
\begin{align}
    \avg{x_{\perp}^2}_{n,J} = \frac{1}{d_{n,J}}\,\frac{\pi^2\,\alpha'}{6}\int d\beta \,e^{\beta\,n}\int\,dk\,e^{-ikJ}e^{\frac{a}{\beta}}\,\beta^{\frac{c}{2}-1}\frac{\frac{\pi\,k}{\beta}}{\sinh\left(\frac{\pi\,k}{\beta}\right)}\,.
\end{align}
This is exactly the same integral one obtains for the expected size for the string in the directions orthogonal to the plane of rotation \eqref{eq:OrthoSize1}.
The sizes obtained from the operator methods and the random walk model will thus match for all regimes of $J$.

\paragraph{Size along the plane of rotation.}

The average size of the random walk in the plane of rotation is given by 
\begin{align}
\label{eq:DefParallelSize}\avg{x_{\parallel}^2}_{n,J} = \frac{1}{d_{n,J}}\int\,dx\,x_{\parallel}^2\,d_{n,J}(x_i)\,,
\end{align}
where $x_{\parallel}$ is either $x_1$ or $x_2$.
Again, evaluating the spatial integral first, we find that 
\begin{align}
\label{eq:x1int}
  \avg{x_{\parallel}^2}_{n,J} =  \frac{\alpha'}{2\,d_{n,J}}\,\int d\beta \,e^{\beta\,n}\int\,dk\,e^{-ikJ}e^{\frac{a}{\beta}}\,\beta^{\frac{c}{2}}\frac{\frac{\pi\,k}{\beta}}{\sinh\left(\frac{\pi\,k}{\beta}\right)}\Bigg[\frac{\pi^2}{\beta}\rb\left(\frac{\pi\,k}{\beta}\right)\Bigg]\,.
\end{align}
One observes that by substituting $k = i\,\beta\,\Omega$ in the integrand one exactly reproduces \eqref{eq:parallelsize}, which was obtained using operator methods.%
\footnote{There is a relative factor of 1/2 between \eqref{eq:x1int} and \eqref{eq:parallelsize} arising because the latter effectively calculates $\avg{x_1^2 + x_2^2}$, which is twice the value given in \eqref{eq:x1int}.}
Therefore, evaluating  $\avg{x_{\parallel}^2}_{n,J}$ in the random walk model exactly reproduces the expected value of the sizes of rotating strings in the plane of rotation \eqref{rparallel}.

By thinking of $n$ as roughly the square of the total length and $|J|$ as the area enclosed by a collection of random walks, a nice geometrical interpretation of the sizes \eqref{rperp} and \eqref{rparallel} emerges:

\begin{itemize}
    \item For $|J|\ll \sqrt{n}$, the total length is much greater than the area and so the random walk part dominates, which is why the sizes in all directions scale to leading order as if there is no rotation present: The square of the sizes are still proportional to the total length.
    \item When $|J|\sim \sqrt{n}$, the size (squared) in the plane of rotation is proportional to the area $|J|$, which now dominates over the random walk part. The remaining budget for random walks is roughly $n-|J|$, which can be interpreted as the total length that is not being used to generate the area in the plane of rotation. This is the reason why the size orthogonal to the plane of rotation is proportional to $\sqrt{n-|J|}$.\footnote{The first correction to the size in the plane of rotation at large values of $|J|$ is also proportional to $\sqrt{n-|J|}$.} 
    \item When $|J|\sim n$ the two-dimensional area takes its maximal value
    since it is equal to the length square up to order-one factors. This provides a random-walk picture of the Regge bound. 
\end{itemize}

%

%

\section*{Acknowledgements}

We thank Iosif Bena, Aditya Hebbar, Jean-Marc Luck, and Yoav Zigdon for useful conversations. 
The work of N\v{C} is supported in part by the Science Foundation Ireland under the grant agreement 22/EPSRC/3832.
RE is supported by MICINN grant PID2022-136224NB-C22, AGAUR grant 2021 SGR 00872, and State
Research Agency of MICINN through the ``Unit of Excellence María de Maeztu 2020-2023'' award to the Institute of Cosmos Sciences (CEX2019-000918-M).
AP and MT are supported by the European Research Council (ERC) under the European Union’s
Horizon 2020 research and innovation programme (grant agreement No 852386). MT is also supported by the Emmy Noether Fellowship program at the Perimeter Institute for Theoretical Physics. 

\appendix
 
\section{The rhotation function 
\texorpdfstring{$\rho(\beta,\Omega)$}{rho}}
\label{app:rhotation}

The average size in the direction along the rotation plane is obtained from the trace \eqref{Rparallel_Rperp} over states with excitation level $N$ and angular momentum $J$, 
\begin{equation}
    \langle R_\parallel^2\rangle \equiv\Tr(R_\parallel^2\, x^{N-\Omega J}).
\end{equation}
To evaluate it we introduce the matrix $x^{N-\Omega J} y^{R_{\parallel}^2}$,
whose trace yields
\begin{equation}
\Tr(x^N y^{R_\parallel^2})=\prod_{n=1}^\infty  \left(\frac{1}{1-x^n}\right)^{D-4}  \frac{1}{1-x^{n+\Omega}y^{\frac{1}{n}}} \frac{1}{1-x^{n-\Omega}y^{\frac{1}{n}}}.
\end{equation}
Differentiating with respect to $y$ and setting $y=1$ gives
\begin{equation}
\label{averageR2parallel}
\badat{2}
       \langle R_\parallel^2\rangle &=\frac{d}{dy}\Tr(x^N y^{R_{\parallel}^2})\big|_{y=1}\\
       &= Z(x,\Omega) \sum_{n=1}^\infty \left(\frac{x^{n+\Omega}}{n(1 - x^{n+\Omega})} + \frac{x^{n-\Omega}}{n(1 - x^{n-\Omega})}\right)\\
       &\equiv \frac{1}{\beta}Z(\beta,\Omega) \rho(\beta,\Omega),
\eadat
\end{equation}
where $x = e^{-\beta}$ and where we defined
\begin{equation}
\label{sumforrho}
    \rho(\beta,\Omega)\equiv\beta \left(\frac{e^{-\beta(n+\Omega)}}{n(1 - e^{-\beta(n+\Omega)})} + \frac{e^{-\beta(n-\Omega)}}{n(1 - e^{-\beta(n-\Omega)})}\right) .
\end{equation}
We want to evaluate this expression in the high-temperature limit where
\begin{equation}
\label{smallbetasumforrho}
    \rho(\beta\approx 0,\Omega) =     \sum_{n=1}^\infty \frac{2}{(n+\Omega)(n - \Omega)}-\beta \sum_{n=1}^\infty\left[ \frac{2}{(n+\Omega)(n - \Omega)}+ \frac{1}{n}\right]
    +\order{\beta^2}     .
\end{equation}
The leading term as $\beta\to 0$ is finite and given by
\begin{equation}
\label{rhosumsmallbeta}
    \rho(\beta\approx 0,\Omega) = \frac{1 - \pi \Omega \cot{(\pi \Omega)}}{\Omega^2}+\order{\beta}\,,
\end{equation}
and we refer to it as the {\it rhotation function}. The $\order{\beta}$ term in \eqref{smallbetasumforrho}, which gives an $\order{\beta^0}$ contribution to the size \eqref{averageR2parallel}, is actually divergent. The same harmonic sum arises already in the computation of the size for the static string \cite{Mitchell:1987th} for which \eqref{sumforrho} reduces to
\begin{equation}
\label{rhostaticlimit}
    \rho(\beta,\Omega=0)=\frac{2}{c}\beta \log Z(\beta),
\end{equation}
where
\begin{equation}
\badat{2}
 \frac{2}{c}\beta \log Z(\beta)
  &= \frac{2}{c}\beta \sum_{n=1}^\infty \frac{e^{-\beta n}}{n(1-e^{-\beta n})}\\
  &\stackrel{\beta\to 0}{=}  \frac{2}{c} \left[\sum_{n=1}^\infty \frac{1}{n^2}- \frac{\beta}{2}\sum_{n=1}^\infty \left(\frac{1}{n^2}+\frac{1}{n}\right)%
  +\order{\beta^2} \right].
\eadat
\end{equation}
While $\sum_{n=1}^\infty \frac{1}{n^2}=\frac{\pi^2}{6}$, the harmonic sum $\sum_{n=1}^\infty\frac{1}{n}$ is divergent. 
In \cite{Mitchell:1987th} the infinite sum in $\log Z(\beta)$ was bounded from above and below. These bounds lead to the asymptotic expression
\begin{equation}
\label{rhoLOandNLO}
     \rho(\beta\approx 0,\Omega=0)= \frac{\pi^2}{3} +\beta \log{\beta}+\order{1}.
\end{equation}
In the presence of rotation we expect
\begin{equation}
\label{rhosumsmallbetaLOandNLO}
    \rho(\beta\approx 0,\Omega) = \frac{1 - \pi \Omega \cot{(\pi \Omega)}}{\Omega^2}+\beta \log \beta+\order{1}\,.
\end{equation}
This is supported by the fact that %
the $\Omega\to0$ limit of $\rho(\beta,\Omega)$ agrees with its static counterpart $\beta\log Z(\beta)$ via \eqref{rhostaticlimit}, and both expressions yield, in the limit that $\beta\to 0$, the same divergent $\Omega$-independent sum at $\order{\beta}$.
In any case, in our computation of string sizes only the rhotation function \eqref{rhosumsmallbeta} will be required, since the subleading corrections are additive terms that are suppressed.

\section{From \texorpdfstring{$\Omega$}{omega} to \texorpdfstring{$J$}{J}}
\label{app:Fourier}
To get the density of states and sizes of the string as a function of the angular momentum $J$ rather than the angular velocity $\Omega$, we compute the Fourier transform $\int dk e^{ikJ} I(k)$ where $k=i\beta \Omega$. For the computation of the density of states in Section~\ref{sec:RotatingStringDensity} the $\Omega$-dependence of the partition function \eqref{dosrot} involves $\frac{\Omega}{\sin(\pi \Omega)}\equiv I$. Its Fourier transform was computed in \cite{Russo:1994ev}
\begin{align}
\label{Fourierdosperpsize}
  \int dk \,e^{-ikJ}\,I(k)= \int dk \,e^{-ikJ}\,\frac{\frac{k}{\beta}}{\sinh\left(\frac{\pi\,k}{\beta}\right)}= \frac{\beta}{2\cosh^2\left(\frac{\beta\,|J|}{2}\right)}\,.
\end{align}
In Section~\ref{ssec:RotatingStringSize} we compute the string sizes orthogonal to and along the plane of rotation. For the former we need again the result \eqref{Fourierdosperpsize}. For the latter we need to compute the Fourier transform of $\frac{\Omega}{\sin(\pi \Omega)} \frac{1 - \pi \Omega \cot{(\pi \Omega)}}{\Omega^2}\equiv I$ which includes also the rhotation function \eqref{rhosumsmallbeta}. We find
\begin{align}
\label{eq:G2IntegralEval}
    \int_{-\infty}^{\infty}\,dk\,e^{-ikJ}\left[- \frac{1- \frac{k\pi}{\beta}\,\coth\left(\frac{k\pi}{\beta}\right)}{\frac{k}{\beta}\,\sinh\left(\frac{k\,\pi}{\beta}\right)}\right] = 2\beta\left[\frac{\beta\aJ}{1+ e^{\beta \aJ}} + \log\left(1+ e^{-\beta\,\aJ}\right)\right]\,.
\end{align}
To obtain this result, we note that the integrand has poles along the imaginary axis at $k = i \beta n$ for $n \in \mathbb{Z}\setminus\{0\}$. 
Then we solve the corresponding contour integral where, depending on the sign of $J$, we need to close the contour as follows.
\begin{itemize}
    \item For $J<0$ and $\mathrm{Im}\;k >0$ we close the contour counter-clockwise in the upper half-plane.
    \item For $J>0$ and $\mathrm{Im}\;k <0$ we close the contour clockwise in the lower half-plane. This gives an extra minus sign when evaluating the integral.
\end{itemize}
Expanding the integrand along the poles we find, for $J>0$, that the sum over residues gives
\begin{align}
    -2\pi i\sum_{\rm residues} = - 2\pi i\sum_{n=1}^{\infty}\frac{i(-1)^{n+1} e^{-n\beta J}\beta(1+ n\beta J)}{n\pi} = 2\beta\left[\frac{\beta J}{1+ e^{\beta J}} + \log\left(1+ e^{-\beta J}\right)\right]\,,
\end{align}
 while for $J<0$ we get
\begin{align}
    2\pi i\sum_{\rm residues} =  2\pi i\sum_{n=1}^{\infty}\frac{i(-1)^{n+1} e^{n\beta J}\beta(-1+ n\beta J)}{n\pi} = 2\beta\left[\frac{-\beta J}{1+ e^{-\beta J}} + \log\left(1+ e^{\beta J}\right)\right].
\end{align}
More compactly this yields \eqref{eq:G2IntegralEval}.

\section{The size of rotating strings}
\label{app:Sizes}

We collect here the necessary formulae to arrive at the expressions \eqref{rperp} and \eqref{rparallel} for the average size of the string transverse to and along the plane of rotation,
\begin{equation} 
\label{appsizes}
\langle \bar{r}^2_\perp \rangle_n \propto \frac{R_{n,\perp}^2}{d_{n,J}},\qquad
\langle \bar{r}^2_\parallel \rangle_n \propto \frac{R_{n,\parallel}^2}{d_{n,J}}.
\end{equation}
The density of states and the size in the direction transverse to and along the rotation plane are
\begin{equation}
\label{dRperpRpara}
    d_{n,J}=\text{const.}\int d\beta e^{n f_d(\beta)}\,, \quad 
    R_{n,\perp}^2=\text{const.}\int d\beta e^{n f_\perp(\beta)} \,,\quad R_{n,\parallel}^2=\text{const.}\int d\beta e^{n f_\parallel(\beta)}\,, 
\end{equation}
where the exponents are given by
\begin{subequations}
    \begin{align}
     n f_d(\beta)&= n \beta+\frac{a}{\beta}+(b+1)\log \beta +\log \big[\sech^2\big(\frac{\beta|J|}{2}\big)\big]\,, \\
     n f_\perp(\beta)&=  n \beta+\frac{a}{\beta}+b\log \beta +\log \big[\sech^2\big(\frac{\beta|J|}{2}\big)\big]\,,\\
     n f_\parallel(\beta)&=  n \beta+\frac{a}{\beta}+b\log \beta +\log \Big[\log(1+e^{-\beta|J|})+\frac{\beta|J|}{1+e^{\beta|J|}}\Big].
    \end{align}
\end{subequations}
The saddle points $\beta_*^d$, $\beta_*^\perp$, $\beta_*^\parallel$ of \eqref{dRperpRpara} are the solutions to $f'_d(\beta_*^d)=0$, $f'_\perp(\beta_*^{\perp})=0$ and $f'_\parallel(\beta_*^{\parallel})=0$, where
\begin{subequations}
    \begin{align}
     n f'_d(\beta)&= n -\frac{a}{\beta^2}+\frac{b+1}{\beta} -\mu_d|J| \,, \\
    n f'_\perp(\beta)&=  n -\frac{a}{\beta^2}+\frac{b}{\beta} - \mu_\perp|J| \,,\\
    n f'_\parallel(\beta)&=  n -\frac{a}{\beta^2}+\frac{b}{\beta} -\mu_\parallel|J|,
    \end{align}
\end{subequations}
and we introduced the shorthand 
\begin{equation}
    \mu_d=\tanh{\left(\frac{\beta |J|}{2}\right)},\qquad \mu_\perp=\tanh{\left(\frac{\beta |J|}{2}\right)},\qquad \mu_\parallel=\frac{\frac{\beta |J|}{4} \;\text{sech}^2{\left(\frac{\beta|J|}{2}\right)}}{\log{(1 + e^{-\beta|J|}) + \frac{\beta|J|}{1 + e^{\beta|J|}}}},
\end{equation}
which only depend on the combination $\beta |J|$.
In the high-temperature limit $\beta\to 0$ we can solve the saddle point equations for the regimes where $\beta |J|$ tends to $0$, a constant value, and $\infty$, for which $\mu_{d,\perp,\parallel}$ is respectively $0$, $\in (0,1)$, and $1$. This has the solution
\begin{subequations}
    \begin{align}
        \beta_*^{d} &= \frac{\sqrt{4a(n-\mu_d|J|)+(b+1)^2}-(b+1)}{2(n-\mu_d|J|)}
        ,\\
        \beta_{*}^{\perp} &= \frac{\sqrt{4a(n-\mu_\perp|J|)+b^2}-b}{2(n-\mu_\perp|J|)}\\
        \beta_{*}^{\parallel} &= \frac{\sqrt{4a(n-\mu_\parallel|J|)+b^2}-b}{2(n-\mu_\parallel|J|)}
        .
    \end{align}
\end{subequations}
The regimes $\beta|J|\to \{0,{\rm const},\infty\}$ are thus consistent with the scaling of the angular momentum $J=\{\order{1},\order{\sqrt{n}}, \order{n}\}$ at large excitation level $n$.\footnote{Note that the regime $\beta|J|\to 0$, which yields saddle point $\beta_*=\sqrt{\frac{a}{n}}$, only captures the leading large $n$ behavior of the slowly rotating string which is independent of $J$ and thus the same as the static case; to capture how the size of slowly rotating strings depends on $|J|$ we will carry out a more refined analysis below.}

\subsection{Size transverse to the rotation plane}
The size at fixed excitation level $n$ transverse to the rotation plane is obtained from
\begin{equation} 
\label{rbar2perp*}
\langle \bar{r}^2_\perp \rangle_n \propto \left(\frac{f''_\perp(\beta_*^{\perp})}
{f''_d( \beta_*^{d})}\right)^{-1/2} e^{n[f_{\perp}(\beta_*^{\perp})-
f_d(\beta_*^{d})]}\,,
\end{equation}
where 
\begin{equation}
\left(\frac{f''_\perp(\beta_*^{\perp})}
{f''_d( \beta_*^{d})}\right)^{-1}=\frac{\frac{2a}{(\beta_*^d)^3}-\frac{b+1}{(\beta_*^d)^2}-\frac{J^2}{2}\sech^2(\frac{\beta_*^d |J|}{2})}{\frac{2a}{(\beta_*^\perp)^3}-\frac{b}{(\beta_*^\perp)^2}-\frac{J^2}{2}\sech^2(\frac{\beta_*^\perp |J|}{2})}
\end{equation}
and
\begin{equation} 
e^{n[f_{\perp}(\beta_*^{\perp})-
f_d(\beta_*^{d})]}=  e^{(\beta_*^{\perp}-
\beta_*^{d})\big[n-\frac{a}{\beta_*^\perp \beta_*^d}\big]}
\left(\frac{\beta_*^\perp}{\beta_*^d}\right)^b\frac{1}{\beta_*^d} \frac{\frac{e^{\beta_*^{\perp}|J|}}{(1+e^{\beta_*^{\perp}|J|})^2}}{\frac{e^{\beta_*^d|J|}}{(1+e^{\beta_*^d|J|})^2}}.
\end{equation}
These expressions drastically simplify since  the saddle points $\beta_*^d$ and $\beta_*^\perp$ agree at leading order in large~$n$, and \eqref{rbar2perp*} is determined by $1/\beta_*^d$.
This yields the perpendicular size $\langle \bar{r}^2_\perp \rangle_n$ as in \eqref{rperp}.

\subsection{Size along the rotation plane} 
The size at fixed excitation level $n$ in the directions along the rotation plane is obtained from
\begin{equation} 
\label{rbar2parallel*}
\langle \bar{r}^2_\parallel \rangle_n \propto\left(\frac{f''_\parallel(\beta_*^{\parallel})}
{f''_d( \beta_*^{d})}\right)^{-1/2} e^{n[f_\parallel(\beta_*^{\parallel})-
f_d(\beta_*^{d})]},
\end{equation}
where 
\begin{align}
&\left(\frac{f''_\parallel(\beta_*^\parallel)}{f''_{d}(\beta_*^{d})}\right)^{-1}=\nonumber\\
    &\frac{\frac{2a}{(\beta_*^d)^3}-\frac{b+1}{(\beta_*^d)^2}-\frac{J^2}{2}\sech^2(\frac{\beta_*^d |J|}{2})}{\frac{2a}{(\beta_*^\parallel)^3}-\frac{b}{(\beta_*^\parallel)^2}-\frac{\frac{J^2}{4}\sech^2(\frac{\beta_*^\parallel |J|}{2})}{\log(1+e^{-\beta_*^\parallel|J|})+\frac{\beta_*^\parallel|J|}{1+e^{\beta_*^\parallel|J|}}}\left(1-\beta_*^\parallel|J| \tanh{\frac{\beta_*^\parallel|J|}{2}}+\frac{(\beta_*^\parallel)^2\frac{|J|^2}{4}\sech^2(\frac{\beta_*^\parallel |J|}{2})}{\log(1+e^{-\beta_*^\parallel|J|})+\frac{\beta_*^\parallel|J|}{1+e^{\beta_*^\parallel|J|}}}\right)}
\end{align}
and
\begin{equation} 
\label{paraexp}
e^{n[f_\parallel(\beta_*^{\parallel})-
f_d(\beta_*^{d})]}\propto e^{(\beta_*^{\parallel}-
\beta_*^{d})\big[n-\frac{a}{\beta_*^\parallel \beta_*^d}\big]}
\left(\frac{\beta_*^\parallel}{\beta_*^d}\right)^b\frac{1}{\beta_*^d} \frac{\log(1+e^{-\beta_*^\parallel|J|})+\frac{\beta_*^\parallel|J|}{1+e^{\beta_*^\parallel|J|}}}{4\frac{e^{\beta_*^d|J|}}{(1+e^{\beta_*^d|J|})^2}}\,.
\end{equation}
At leading order we have
\begin{equation}
    \beta_{*}^{\parallel}-\beta_{*}^{d}=\begin{cases}
        \frac{1}{2n}+\order{n^{-3/2}}, &J=\order{1}\\ \frac{1}{2n}\left(1+(\mu_{\parallel}-\mu_d)\sqrt{\frac{a}{n}}|J|\right)+\order{n^{-3/2}},&J=\order{\sqrt{n}}\\\frac{1}{2(n-|J|)}+\order{n^{-3/2}},&J=\order{n}
    \end{cases},
\end{equation}
where we note that $\mu_{\parallel}>\mu_d$, and
\begin{equation}
\beta_{*}^{\parallel}\beta_{*}^{d}=\begin{cases}
        \frac{a}{n}+\order{n^{-3/2}},&J=\order{1}\\ \frac{a}{n}+\order{n^{-3/2}}%
        ,&J=\order{\sqrt{n}}\\\frac{a}{n-|J|}+\order{n^{-3/2}},&J=\order{n}
    \end{cases}.
\end{equation}
It follows that the exponent in \eqref{rbar2parallel*} is constant at leading order. 
The same goes for 
\begin{equation}
    \frac{\beta_{*}^{\parallel}}{\beta_{*}^{d}}=1+\order{n^{-1/2}}
    ,
\end{equation}
whose leading order term is 1 for all three cases $J=\order{1},\order{\sqrt{n}}, \order{n}$ and, similarly,
\begin{equation}
   \sqrt{\frac{f''_d(\beta_*^d)}{f''_{\parallel}(\beta_*^{\parallel})}}%
   =1+\order{n^{-1/2}}.
\end{equation}
The last term in \eqref{paraexp} scales as 
\begin{equation}
    \frac{\log(1+e^{-\beta_*^\parallel|J|})+\frac{\beta_*^\parallel|J|}{1+e^{\beta_*^\parallel|J|}}}{4\frac{e^{\beta_*^d|J|}}{(1+e^{\beta_*^d|J|})^2}}=\begin{cases}
        \log (2),&J=\order{1}\\ \log (2)\,C_\parallel\left(\frac{|J|}{\sqrt{n}}\right),&J=\order{\sqrt{n}}\\ \frac{\sqrt{a}|J|}{4\sqrt{n-|J|}},&J=\order{n}
    \end{cases},
\end{equation}
where $C_\parallel(J/\sqrt{n})$ is given in \eqref{Ceq}.
Thus the last two terms in \eqref{paraexp} together behave as %
 $\sqrt{n}$, $C_\parallel(J/\sqrt{n})\sqrt{n-\mu_d|J|}$, and $|J|$  for, respectively, $J=\order{1}$, $\order{\sqrt{n}}$, $\order{n}$.
 From here we obtain the parallel size $\langle \bar{r}^2_\parallel \rangle_n$ as in \eqref{rparallel}.

\subsection{Overall factors and the ratio of sizes}\label{subsec:overratio}

So far we have ignored all overall $\order{1}$ constant prefactors of ${\langle \bar{r}^2_\perp \rangle_n}$ and $ {\langle \bar{r}^2_\parallel \rangle_n}$, 
since the black hole/string correspondence is insensitive to them.
However, if we include them an apparent puzzle appears when we send  $J\to 0$.
In this limit, the string state becomes spherical on average and the distinction between the two sizes, along the rotation plane and transverse to it, should disappear. That is, the ratio between the two sizes that we study in the main text should naturally become one when $J\to 0$.

Instead, the results of our calculations for $\langle \bar{r}^2_{\parallel}\rangle_n$ and $\langle \bar{r}^2_\perp \rangle_n$ \emph{do not} become the same when $J\to 0$.
We find that\footnote{For the comparison of the two sizes, here we are considering  $\langle \bar{r}^2_\parallel \rangle=\langle X_1^2+X_2^2\rangle/2$ and $\langle \bar{r}^2_\perp \rangle=\langle X_3^2+\dots+X_{c}^2\rangle/(c-2)$.}
\begin{equation}\label{discrep}
    \langle \bar{r}^2_\parallel \rangle_{n,J=0} =2\log2\simeq 1.39\,,\qquad
    \langle \bar{r}^2_\perp \rangle_{n,J=0} = \frac{\pi^2}{6}\simeq 1.65\,.
\end{equation}
The discrepancy may look troublesome. However, its origin is easy and instructive to track down.
The two radii in \eqref{discrep} are the limiting results of two different calculations: along the plane of rotation, the value of the spin is first fixed to be exactly $J$, and then sent to exactly $J=0$. In the transverse directions, instead, the spin is allowed to fluctuate, and only its average is required to vanish. The latter calculation allows for less constrained fluctuations than the former: configurations with spin $+J$ and $-J$ are both considered, and including them yields a larger average size, as we find in  \eqref{discrep}.

For the purposes of our discussion in the main text we can ignore this effect, since our focus is on the spin dependence of the size ratio and not on any numerical factors. Accounting for this effect also justifies why, as we discuss after \eqref{modratio}, the fact that $C_\parallel>1$ is a meaningful indication of a larger spread along the rotation plane than in the transverse directions.

Finally, note that in the black hole side there is no similar discrepancy in the two sizes since the calculation does not involve any averaging.

\section{Statistical properties of closed fundamental strings}
\label{sec:detsrot}
\label{sec:closed}

Building on results from the open string partition function, one can show that the partition function for a closed rotating string is just a square of the open string partition functions 
\begin{equation}
 Z_{\text{closed}}(\beta, \Omega_L, \Omega_R) = Z_{\text{open}}(\beta, \Omega_L) \times Z_{\text{open}}(\beta, \Omega_R), \label{closedPF}
\end{equation}
with level matching imposed, where $Z_{\text{open}}$ is the partition function for the left- and right-moving sector, which behaves like an open string with angular velocity $\Omega_{L, R}$. 
We need to obtain the density of states and the relevant sizes, and both depend on $\Omega_L$ and $\Omega_R$. Let us look at the density of states first. We can write
\begin{equation}
\label{eq:ClosedNumber1}
    d_{n, J_L, J_R} = -\frac{1}{2\pi i} \int d\beta\; e^{n\beta} \int_{-\infty}^\infty \frac{dk_L}{2\pi} e^{-i k_L J_L} Z(\beta, k_L) \int_{-\infty}^\infty \frac{dk_R}{2\pi} e^{-i k_R J_R} Z(\beta, k_R), 
\end{equation}
where $Z(\beta, k_i)$ has the same form as a single open string partition function. We can evaluate the Fourier integrals explicitly and obtain a similar form as \eqref{dosrot5}, namely
\begin{equation}
    d_{n, J_L, J_R} = \text{const.} \int d\beta \exp{n\beta + \log{Z(\beta, J_L) + \log{Z(\beta, J_R)}}}
\end{equation}
which gives as a saddle point equation
\begin{equation}
    n + \frac{\tilde{b}+1}{\beta} - \frac{\tilde{a}}{\beta^2} - J_L \tanh{\left(\frac{\beta J_L}{2}\right)} - J_R \tanh{\left(\frac{\beta J_R}{2}\right)} = 0,
\end{equation}
where $\tilde{b} = b_L + b_R + 1$ and $\tilde{a} = a_L + a_R$. Assuming that both $J$'s scale the same with $n$, we obtain similar solutions to the saddle point equation, with 
\begin{equation}
    \beta^d_* = \sqrt{\frac{\tilde{a}}{n - \tilde{\mu}_d |J|}}, \hspace{15pt} \tilde{\mu}_d = c_L \mu_d^L + c_R \mu_d^R, \hspace{10pt} \mu_d^i = \tanh{\left(\frac{\beta c_i |J|}{2}\right)},
\end{equation}
where $i = L, R$ and which, as before, is valid for small and large values of both $J$'s. Note that we have defined $J_i \equiv c_i J$. From here it is clear that we obtain the same form of the result as for the open string case
\begin{equation}
\label{eq:ClosedNumber2}
    d_{n, J_L, J_R} \propto \frac{1}{4}\left(n - \tilde{\mu}_d|J|\right)^{-(\tilde{b}+1)/2 - 3/4} e^{\frac{\sqrt{\tilde{a}} (2n - \tilde{\mu}_d|J|)}{\sqrt{n - \tilde{\mu}_d|J|}}}\text{sech}^2\left(\frac{\sqrt{\tilde{a}} J_L}{2 \sqrt{n - \tilde{\mu}_d|J|}}\right)\text{sech}^2\left(\frac{\sqrt{\tilde{a}} J_R}{2 \sqrt{n - \tilde{\mu}_d|J|}}\right).
\end{equation}
For large $J$'s, this density of states can look nicer, as was discussed in \cite{Russo:1994ev}. For typical values of $J_L$ and $J_R$, we assume $\beta J_L \sim \beta J_R \sim \order{1}$ and so the saddle point is given by
\begin{equation}
\label{eq:ClosedBetaStar}
    \beta^d_* = \frac{\sqrt{4 \tilde{a} (n-\tilde{\mu}_d  |J|)+(\tilde{b} + 1)^2}-(\tilde{b}+1)}{2(n- \tilde{\mu}_d |J|)},
\end{equation}
where $\tilde{\mu}_d$ is a constant in the range $\tilde{\mu}_d \in (0,1)$.

\paragraph{Size transverse to the rotation plane.}

Similarly to the open string case, we can write
\begin{equation}
    R^2_\perp = \sum_{n=1}^\infty \frac{(a_L)^\dagger_n (a_L)_n}{n} + \sum_{m=1}^\infty \frac{(a_R)^\dagger_m (a_R)_m}{m} = R^L_\perp + R^R_\perp,  
\end{equation}
where we have chosen one of the directions $i \neq 1,2$. To calculate the size, we need to evaluate 
\begin{equation}
\begin{split}
    \text{tr}\left(x^{N + \bar{N}} y^{R^L_\perp + R^R_\perp}\right) &= 
    \prod_n \left(\frac{1}{1-x^n}\right)^{D-1} \frac{1}{1 + x^{n+\Omega_L}} \frac{1}{1 + x^{n - \Omega_L}} \frac{1}{1 - x^n y^{1/n}} \\ &\times \prod_m \left(\frac{1}{1-x^m}\right)^{D-1} \frac{1}{1 + x^{m+\Omega_R}} \frac{1}{1 + x^{m - \Omega_R}} \frac{1}{1 - x^m y^{1/m}},
\end{split}
\end{equation}
where we will use
\begin{equation}
    \frac{d}{dy} \left(\prod_{n=1}^\infty f_n(y) \right) =  \sum_{n=1}^\infty f'_n(y) \prod_{n \neq k}^\infty f_k(y), \label{product}
\end{equation}
where the product runs over all the indices different than $n$, and $1 \leq k \leq \infty$. Since we need to set $y = 1$, we see that the new terms will just be added as open string partition functions with appropriate angular velocities. Following the same logic as for the open string perpendicular size, we obtain
\begin{equation}
\label{eq:ClosedPerpInt}
    R_{n,\perp}^2 = \frac{2}{c} Z_{closed} (x, \Omega_L, \Omega_R) \log Z(x),
\end{equation}
where $c = D - 2$, and $Z(x)$ is the partition function of a static, open string. The saddle point equation becomes
\begin{equation}
    n f_\perp = n\beta  + \tilde{b} \log{\beta} + \frac{\tilde{a}}{\beta} + 2 \log\sech\left(\frac{\beta |J_L|}{2}\right) + 2 \log\sech\left(\frac{\beta |J_R|}{2}\right).
\end{equation}
We can now use this expression to obtain the perpendicular size of the closed string. To do this, we use the same method as for the open string, and since the expression is very similar to \eqref{perpsize}, we can immediately extract the size for all three limits of $J_L$ and $J_R$ (always assuming that they scale in the same way),
\begin{equation}
    \expval{\bar{r}^2_\perp}_n = \frac{R^2_{n,\perp}}{d_{n, J_L, J_R}} = \frac{2}{c} \log Z(\beta^d_*), 
\end{equation}
where $\beta_*^d$ is given by \eqref{eq:ClosedBetaStar} at leading order in $n$ (for subleading order, we change $\tilde{b} + 1 \to \tilde{b}$). Depending on the value of $\tilde{\mu}_d$ and $\tilde{b}$, we obtain all three limits of $J_L$ and $J_R$. Hence, we obtain
\begin{equation}
\label{eq:PerpSizeClosed}
    \langle \bar{r}^2_\perp \rangle_n = \begin{cases}
     \sqrt{n}, &  J_L, J_R = \order{1} \\
      \sqrt{n - \tilde{\mu}_d|J|}, &  J_L, J_R = \order{\sqrt{n}} \\
      \sqrt{n- \kappa|J|}, & J_L, J_R = \order{n}
    \end{cases}
\end{equation}
where again $\tilde{\mu}_d\in (0,1)$ and $\mu_d^i = 1$ for large angular momenta, and $\kappa = c_L + c_R$.

\paragraph{Size along the rotation plane.}

As usual, the parallel size calculation will require more effort, but not more than the open string case. We define
\begin{equation}
    R^2_\parallel = \sum_n \frac{1}{n}\left((a_L)^\dagger_n (a_L)_n + (b_L)^\dagger_n (b_L)_n\right) + \sum_m \frac{1}{m}\left((a_R)^\dagger_m (a_R)_m + (b_R)^\dagger_m (b_R)_m\right).
\end{equation}
The trace will give us
\begin{equation}
    \prod_n \frac{1}{(1 - x^n)^{D-2}} \frac{1}{1 - x^{n-\Omega_L} y^{1/n}} \frac{1}{1 - x^{n+\Omega_L} y^{1/n}} \times  \prod_m \frac{1}{(1 - x^m)^{D-2}} \frac{1}{1 - x^{m-\Omega_R} y^{1/m}} \frac{1}{1 - x^{m+\Omega_R} y^{1/m}},
\end{equation}
and since we again have the same form as for the perpendicular case, we obtain 
\begin{equation}
    \langle R_{\parallel}^2 \rangle = \sum_{n=1}^\infty \sum_{\mu = L}^R \left(\frac{x^{n + \Omega_\mu}}{n(1 + x^{n+\Omega_\mu})} + \frac{x^{n - \Omega_\mu}}{n(1 - x^{n-\Omega_\mu})}\right) Z_{closed} (x, \Omega_L, \Omega_R).
\end{equation}
To extract the size, we need to first Fourier transform to obtain expressions in terms of angular momenta, and then perform the saddle point approximation for the $x$-integral,
\begin{equation}
\label{eq:InvLapArgument}
    \int \frac{dk_L}{2\pi} e^{-i k_L J_L} \int \frac{dk_R}{2\pi} e^{-i k_R J_R}  \langle R_{\parallel}^2 \rangle = \beta^{\tilde{b}} e^{\tilde{a}/\beta} \left(\frac{\log{(1 + e^{-J_L \beta})} + \frac{J_L \beta}{1 + e^{J_L \beta}}}{\cosh^2(\beta J_R/2)} + J_L \leftrightarrow J_R\right).
\end{equation}
We see that we will have two integrals over $x$, and we also see that the logarithm will separate the fraction above and give us the same terms that we already had. More explicitly,
\begin{equation}
\begin{split}
\label{eq:ParallelSaddleClosed}
    \int d\beta \exp{n\beta + \tilde{b}\log \beta + \frac{\tilde{a}}{\beta} + \log{\left(\log{(1 + e^{-|J_L| \beta})} + \frac{|J_L| \beta}{1 + e^{|J_L| \beta}}\right)} -  \log{\cosh^2\left(\frac{\beta |J_R|}{2}\right)} } + J_L \leftrightarrow J_R.
\end{split}
\end{equation}
In order words, we will split
\begin{equation}
    R^2_{n, \parallel} = R^2_L + R^2_R,
\end{equation}
where
\begin{equation}
    R^2_L = \int d\beta \exp\left\{n\beta + \tilde{b}\log \beta + \frac{\tilde{a}}{\beta} + \log{\left(\log{(1 + e^{-|J_L| \beta})} + \frac{|J_L| \beta}{1 + e^{|J_L| \beta}}\right)} +  2\log{\sech\left(\frac{\beta |J_R|}{2}\right)} \right\}
\end{equation}
and 
\begin{equation}
    R^2_R = \int d\beta \exp\left\{n\beta + \tilde{b}\log \beta + \frac{\tilde{a}}{\beta} + \log{\left(\log{(1 + e^{-|J_R| \beta})} + \frac{|J_R| \beta}{1 + e^{|J_R| \beta}}\right)} +  2\log{\sech\left(\frac{\beta |J_L|}{2}\right)} \right\}
\end{equation}
The saddle points are now easily obtained (and will be equivalent at leading order in $n$) since both of the $J$-dependent terms have similar behavior in all three limits of interest (again, assuming $J_L$ and $J_R$ scale in the same way with $n$). Throughout this calculation, we will heavily rely on previous calculations, especially from Appendix~\ref{app:Sizes}. The ratios of one-loop factors will be equal to one in all cases, given that the first part of the saddle point equations will be the same as for the open string parallel size calculation. Additionally, the second part will cancel with the density of states since they scale in the same way.\footnote{There will be a factor of two difference coming from the fact that the density of states involves both $J_L$ and $J_R$, while in the saddle, only one of them will feature; at our level of precision, this does not matter.}

In what follows, the one-loop factors always go to 1, so we will not write them. The relevant saddle point equations are given by
\begin{equation}
    n f_i = n\beta  + \tilde{b} \log{\beta} + \frac{\tilde{a}}{\beta} + \log\left(\log\left(1 + e^{\beta |J_i|}\right) + \frac{\beta |J_i|}{1 + e^{\beta |J_i|}}\right) + 2 \log\sech\left(\frac{\beta |J_j|}{2}\right),
\end{equation}
where $i, j = L, R$. For all of these cases, the saddle point equation becomes
\begin{equation}
    n + \frac{\tilde{b}}{\beta} - \frac{\tilde{a}}{\beta^2} - \tilde{\mu}_\parallel |J| = 0, 
\end{equation}
where $\tilde{\mu}_\parallel = c_L \mu_\parallel^L + c_R \mu_\parallel^R$, and
\begin{equation}
    \mu_\parallel^i = \frac{1}{4}\frac{\beta |J_i| \sech^2{\left(\frac{\beta |J_i|}{2}\right)}}{\log{\left(1 + e^{-\beta |J_i|}\right)} + \frac{\beta |J_i|}{1 + e^{\beta |J_i|}}}, \hspace{10pt} i = L, R,
\end{equation}
keeping in mind that in the large angular momenta limit, we take $\tilde{\mu}_\parallel = 1$. Neglecting the $\tilde b$-term, we obtain 
\begin{equation}
    \beta_*^\parallel = \sqrt{\frac{\tilde{a}}{n - \tilde{\mu}_\parallel|J|}}.\label{betapar}
\end{equation}
We want to calculate 
\begin{equation}
    \frac{R^2_{n, \parallel}}{d_{n, J_L, J_R}} = \left(\frac{\beta_*^\parallel}{\beta_*^d}\right)^{\tilde{b}} \frac{4}{\beta_*^d} e^{n(\beta_*^\parallel - \beta_*^d) + \tilde{a}\left(\frac{1}{\beta_*^\parallel} - \frac{1}{\beta_*^d}\right)} F_\parallel(\beta, J),
\end{equation}
where
\begin{equation}
     F_\parallel(\beta, J) = \frac{g(J_L, \beta_*^\parallel) h(J_R, \beta_*^\parallel) + g(J_R, \beta_*^\parallel) h(J_L, \beta_*^\parallel)}{h(J_L, \beta_*^d) h(J_R, \beta_*^d)}, 
\end{equation}
with 
\begin{equation}
    g(J, \beta) = \log\left(1 + e^{-\beta |J|}\right) + \frac{\beta |J|}{1 + e^{\beta |J|}}, \hspace{15pt}  h(J, \beta) = \sech^2\left(\frac{\beta |J|}{2}\right).
\end{equation}
Notice that for moderate $J$, we can approximate the product $\beta J \sim \order{1}$, so that the function $F_\parallel(\beta, J)$ becomes a constant as it only depends on such a product. We can then write $F_\parallel(\beta |J| \sim \order{1}) \equiv 2\log{(2)} \tilde{C}_\parallel$ where we see from the functional form that this constant is manifestly positive, and we took out a factor of $2\log{(2)}$ to connect with the small $J$ result, as for the open string case. From the other terms, we obtain a non-trivial contribution only from the $1/\beta_*^d$ term, at least to the order $\order{n^{-1/2}}$, which we can then directly read off from \eqref{betapar}. As for the small and large $J$ limit, we have $\beta_*^d = \beta_*^\parallel$ as $\tilde{\mu}_d = \tilde{\mu}_\parallel = (0,1)$. 

Therefore, we can write the function $F_\parallel(\beta, J)$ for all three cases
\begin{equation}
    F_\parallel(\beta, J) = \begin{cases}
        2 \log{(2)} , & J_L, J_R = \order{1} \\
         2 \log{(2)}  \tilde{C}_\parallel\left(J/\sqrt{n}\right), & J_L, J_R = \order{\sqrt{n}} \\
        \frac{|J_L|}{4} \sqrt{\frac{\tilde{a}}{n - |J_L|}} + \frac{|J_R|}{4} \sqrt{\frac{\tilde{a}}{n - |J_R|}}, & J_L, J_R = \order{n}
    \end{cases}
\end{equation}
The final result for the size is then given by the product $1/\beta_*^d F_\parallel(\beta, J)$,
\begin{equation}
    \label{eq:ParSizeClosed}
      \frac{\langle \bar{r}^2_\parallel \rangle_n}{\ell_s^2} \propto \begin{cases}
     \sqrt{n} , &  J_L, J_R = \order{1} \\
    C_\parallel\sqrt{n-\mu_d|J|}, &  J_L, J_R = \order{\sqrt{n}} \\
     |J| , & J_L, J_R = \order{n}
    \end{cases}
\end{equation}
where we neglected higher-order corrections and the overall constants. Since all the functional forms are the same as in the open string case, we reproduce \eqref{eq:ResRatio}.

\paragraph{Sizes beyond the leading order in large excitation level $n$.} Since the correspondence principle works best for small values of angular momenta, we would like to see if the subleading corrections have the same functional form as the open string case, and therefore, the black hole. Our saddle point equation now becomes

\begin{equation}
n + \frac{\tilde{b}}{\beta} + \frac{\tilde{a}}{\beta^2} - \frac{\kappa_i}{2} \beta |J|^2 = 0
\end{equation}
which implies
\begin{equation}
    \beta_*^i = \sqrt{\frac{\tilde{a}}{n}} - \frac{\tilde{b}}{2n} + \frac{\tilde{b}^2}{8\sqrt{\tilde{a}} n^{3/2}} + \frac{\tilde{a} \kappa_i |J|^2}{4 n^2}, 
\end{equation}
where
\begin{equation}
     \kappa_i = \begin{cases}
     c_L^2 + c_R^2, &  \text{for $\perp$ size} \\
      c_L^2 + \frac{c_R^2}{2\log 2}, &  \text{for $\parallel_R$  size} \\
      c_R^2 + \frac{c_L^2}{2 \log 2}, & \text{for $\parallel_L$  size}.
    \end{cases}
\end{equation}
To calculate small corrections for the perpendicular size, we need to evaluate
\begin{equation}
    \frac{R^2_{n,\perp}}{d_{n, J_L, J_R}} = \frac{2\tilde{a}}{c} \,\sqrt{\frac{f_d''(\beta_*^d)}{f_{\perp}''(\beta_*^{\perp})}}\,\left(\frac{\beta_*^\perp}{\beta_*^d}\right)^{\tilde{b}} \frac{1}{\beta_*^d} e^{n(\beta_*^\perp - \beta_*^d) + \tilde{a}\left(\frac{1}{\beta_*^\perp} - \frac{1}{\beta_*^d}\right)} F_\perp(\beta, J), 
\end{equation}
where
\begin{equation}
    F_\perp(\beta, J) = \frac{h(J_L, \beta_*^\perp) h(J_R, \beta_*^\perp)}{h(J_L, \beta_*^d) h(J_R, \beta_*^d)}, \hspace{15pt} h(J, \beta) = \sech^2\left(\frac{\beta |J|}{2}\right).
\end{equation}

For the parallel size small corrections, we have to evaluate separately the left and right contributions,
\begin{equation}
    \frac{R^2_L}{d_{n, J_L, J_R}} = \sqrt{\frac{f_d''(\beta_*^d)}{f_{L}''(\beta_*^{L
})}}\,\left(\frac{\beta_*^L}{\beta_*^d}\right)^{\tilde{b}} \frac{4}{\beta_*^d} e^{n(\beta_*^L - \beta_*^d) + \tilde{a}\left(\frac{1}{\beta_*^L} - \frac{1}{\beta_*^d}\right)} F_L(\beta, J), \label{parnew}
\end{equation}
where
\begin{equation}
    F_L(\beta, J) = \frac{g(J_L, \beta_*^L) h(J_R, \beta_*^L)}{h(J_L, \beta_*^d) h(J_R, \beta_*^d)}, \hspace{15pt} g(J, \beta) = \log\left(1 + e^{-\beta |J|}\right) + \frac{\beta |J|}{1 + e^{\beta |J|}}.
\end{equation}
The same equations hold for the R movers, just with $L \leftrightarrow R$. Let us evaluate these expressions simultaneously: we can do so since the form of the small $J$ saddle point is the same, just with a different $\kappa$ parameter. Note that we cannot neglect the one-loop factors. 
We can therefore write 
\begin{equation}
\frac{\langle \bar{r}^2_\perp\rangle_n}{\ell_s^2} \propto \sqrt{\frac{n}{\tilde{a}}} + \tilde{\gamma}_1 + \frac{\tilde{\gamma}_2}{\sqrt{n}} + \order{n^{-1}}
\end{equation}
and 
\begin{equation}
    \frac{\langle \bar{r}^2_\parallel\rangle_n}{\ell_s^2} \propto \sqrt{\frac{n}{\tilde{a}}} + \tilde{\gamma}_1 + \frac{\tilde{\gamma}_2}{\sqrt{n}} + |J|^2 \frac{\kappa_d \tilde{\gamma}_\parallel}{\sqrt{n}}+ \order{n^{-1}},
\end{equation}
where 
\begin{align}
    &\tilde{\gamma}_1 = \frac{3 + 2\tilde{b}}{4\, \tilde{a}}, && \tilde{\gamma}_2 = \frac{11 + 16\tilde{b}+ 4\tilde{b}^2}{32 \tilde{a}^{3/2}}, && \tilde{\gamma}_\parallel = \sqrt{\tilde{a}} \frac{2\log 2 - 1}{16 \log 2}, && \kappa_d = c_L^2 + c_R^2.
\end{align}
From here we can  obtain the ratio
\begin{equation}
\frac{\langle \bar{r}^2_\perp\rangle_n}{\langle \bar{r}^2_\parallel\rangle_n} = 1 - \kappa_d \sqrt{\tilde{a}} \tilde{\gamma}_\parallel \frac{|J|^2}{n}.
\end{equation}
Since all constants are positive, the overall effect leads to the size shrinkage. The proportionality constant in the perpendicular case is $\frac{2\tilde{a}}{ c}$, while for parallel it is just a factor of $8 \log 2$. We see that all results qualitatively match those of the open string calculations.

\section{Details regarding the random walk model}
\label{app:RandomWalk}

In this appendix we provide some more technical details of the random walk model introduced in Section~\ref{sec:randomwalk}.
In addition, we show how to modify the model to reproduce the sizes of closed strings, as discussed in Appendix~\ref{sec:closed}, and show that this involves considering two independent closed random walks.

\subsection*{Setting up the path integral}
\label{sapp:Setup}
We use the same setup as discussed in Section~\ref{sec:randomwalk}, but we allow the parameter $s$, used to parametrise the random walks $X_i(s) \in \mathbb{R}^c$, to take on values $s \in [0,\ell)$, where $\ell$ is general.
As such, the boundary condition of the random walk starting and ending at a point $x_i \in \mathbb{R}^c$ is modified to 
\begin{align}
\label{eq:ABC}
    X_i(0)= X_i(\ell) = x_i\,,
\end{align}
and we define the properties of the random walks as
\begin{align}
    \label{eq:AProperties}
    n\left[X_i\right]= \frac{1}{4\pi\,\alpha'}\,\int_0^\ell\,\dot{X}_i\,\dot{X}^i\,ds\,,\quad
    J\left[X_i\right] = \frac{1}{2\pi\,\alpha'}\,\int_0^\ell\left(X_1\,\dot X_2 - X_2\,\dot X_1\right)\,ds\,,
    \quad \left\langle X_i\right\rangle = \int_0^\ell X_i\,ds\,.
\end{align}
We are interested in the number of random walks with fixed values of $n[X_i]=n$ and $J=J[X_i]$ that pass through the point $x_i$ and have their center of mass at the origin
\begin{align}
\label{eq:ADensity2}
d_{n,J}(x_i) \propto \sum_{P_c(x_i)}\delta\left(n 
- n\left[X\right]\right) \,\delta\left(J-J\left[X\right]\right)\,\delta\left(\avg{X_i}\right)\,,
\end{align}
where $P_c(x_i)$ are all closed loops passing through point $x_i$. 
Using the definitions \eqref{eq:AProperties}, we can express this sum over closed loops as a path integral%
\footnote{To go from \eqref{eq:ADensity2} to \eqref{eq:APIRot}, we use the following representations of the Dirac deltas
\begin{align*}
    \delta(x_i) = \frac{1}{2\pi}\int_{-\infty}^{\infty}\,d\mu \,e^{-i\,\mu\,x}\,,\qquad \delta(x_i) = \frac{1}{2\pi\,i}\int_{c_0-i\infty}^{c_0+i\infty}\,d\beta \,e^{\beta\,x_i}\,,
\end{align*}
and have reabsorbed the constant prefactors into the normalisation $\cN(\beta,k)$.}
\begin{align}
    \label{eq:APIRot}
    d_{n,J}(x_i) =   \int_{X_i(0) = x_i}^{X_i(\ell) =x_i}\,\cD X_i\,\int\,d\beta \,e^{\beta\,n}\,\int\,dk\,\cN(\beta,k)\,e^{-i\,k\,J}\int\,d\mu_i\,e^{-I_k[X_i]}\,,
\end{align}
where we sum over all paths starting and ending at $x_i$ weighted by the action
\begin{align}
\label{eq:AAction1}
     I_k[X_i] = \frac{1}{{4\pi\,\alpha'}}\int_0^{\ell}\,\left[{\beta}\dot X_i\,\dot X^i+ 4\pi i\,\alpha'\,\mu_i\,X_i- 2i\, k\left(X_1\,\dot X_2 - X_2 \,\dot X_1\right)\right]\,ds\,.
\end{align}
In the path integral, we also introduced a normalisation factor $\cN(\beta,k)$, which for now we keep undetermined, in contrast to the main text.

To evaluate \eqref{eq:PIRot}, we assume that we can exchange the order of integration
\begin{align}
    \label{eq:APIRot2}
    d_{n,J}(x_i) =   \int\,d\beta \,e^{\beta\,n}\,\int\,dk\,e^{-i\,k\,J}\,\cN(\beta,k)\int\,d\mu_i\,\int_{X_i(0) = x_i}^{X_i(\ell) =x_i}\,\cD X_i\,e^{-I_k[X_i]}\,,
\end{align}
and begin by evaluating the path integral while keeping the other parameters fixed.
Since the action \eqref{eq:AAction1} is quadratic, we can 
solve the path integral by finding the on-shell action and the one-loop factor 
\begin{align}
    \int_{X(0) = x_i}^{X(\ell) =x_i}\,\cD X_i\,e^{-I_k[X]} = f(\beta,k)\,e^{-I_k^{(\rm c)}}\,.
\end{align}
The equations of motion are
    \begin{gather}
    \label{eq:AEOMRot}        \beta\,\ddot{X}_1 -2\pi\, i\,\alpha' \,\mu_1 + 2\,i\,k \,\dot{X}_2=0\,,\quad
        \beta\,\ddot{X}_2- 2\pi\, i\,\alpha' \,\mu_2 - 2\,i\,k \,\dot{X}_1=0\,,\quad 
        \beta\,\ddot{X}_a - 2\pi\, i\,\alpha' \,\mu_a=0\,, 
    \end{gather}
where $a= 3,\ldots c$. 
The solutions to these equations, subject to the  boundary conditions \eqref{eq:ABC}, are
\begin{subequations}
    \label{eq:AEOMSolRot}
    \begin{align}
        X_1(s) &= x_1 - \frac{\pi\,\alpha'}{2\,k\,\sinh\left(\frac{\ell\,k}{\beta}\right)}\Bigg[2\,i\,\ell\,\mu_1 \sinh\left(\frac{(\ell-s)k}{\beta}\right)\,\sinh\left(\frac{s\,k}{\beta}\right)\nonumber\\
        &\qquad \hspace{10em}+\mu_2\left(\ell \sinh\left(\frac{(\ell-2s)k}{\beta}\right)- (\ell-2s)\sinh\left(\frac{\ell\,k}{\beta}\right)\right)\Bigg]\,,\\
        X_2(s) &= x_2 - \frac{\pi\,\alpha'}{2\,k\,\sinh\left(\frac{\ell\,k}{\beta}\right)}\Bigg[2\,i\,\ell\,\mu_2 \sinh\left(\frac{(\ell-s)k}{\beta}\right)\,\sinh\left(\frac{s\,k}{\beta}\right)\nonumber\\
        &\qquad \hspace{10em}-\mu_1\left(\ell \sinh\left(\frac{(\ell-2s)k}{\beta}\right)- (\ell-2s)\sinh\left(\frac{\ell\,k}{\beta}\right)\right)\Bigg]\,,\\
        X_a(s) &= x_a - \frac{i\,\pi\,\alpha'\,\mu_a}{\beta}\,s\,(\ell -s)\,.
    \end{align}
\end{subequations}
As a consistency check, setting $k=0$ reduces the solutions for $X_1$ and $X_2$ to the same form as those for the remaining directions $X_a$.
Inserting these into the action yields 
\begin{equation}
    \label{eq:AActionRot}
    I_k^{(\rm c)} = \frac{3\,\beta}{\pi\,\alpha'\,\ell}\left(\frac{x_1^2 + x_2^2}{3\,\rb\left(\frac{\ell\,k}{\beta}\right)}+x_a^2\right) + \frac{\pi\,
    \alpha'\,\ell^3\,\rb\left(\frac{\ell\,k}{\beta}\right)}{4\,\beta}\sum_{i=1,2}\left[\mu_i+ \frac{2\,i\,\beta\,x_i}{\pi\,\alpha'\,\ell^2\,\rb\left(\frac{\ell\,k}{\beta}\right)} \right]^2 +\frac{\pi\,\alpha'\,\ell^3}{12\,\beta}\left[\mu_a+ \frac{6\,i\,\beta\,x_a}{\pi\,\alpha'\,\ell^2}\right]^2\,,
\end{equation}
where we assume a summation over $a$ indices and defined the function
\begin{align}
    \rb(y) \equiv \frac{\coth y}{y}- \frac{1}{y^2}\,.
\end{align}
The one-loop contribution can be determined by demanding that
\begin{align}
    \int_{X_i(0) = x_i}^{X_i(\ell+t) =x_i}\,\cD X_i\,e^{-I_k[X_i]} = \int_{-\infty}^{\infty}\,dy_i \int_{X_i(0) = x_i}^{X_i(\ell) =y_i}\,\cD X_i\,e^{-I_k[X_i]}\int_{X_i(0) = y_i}^{X_i(t) =x_i}\,\cD X_i\,e^{-I_k[X_i]}\,,
\end{align}
which gives
\begin{align}
\label{eq:AOneLopp}
    f(\beta, k) = \left(\frac{\beta}{4\pi^2\,\alpha'\,\ell}\right)^{\frac{c}{2}}\frac{\frac{\ell\,k}{\beta}}{\sinh\left(\frac{\ell\,k}{\beta}\right)}\,.
\end{align}
Evaluating  this one-loop contribution is not necessary, since it can be absorbed into the normalisation $\cN(\beta,k)$. 
However, it is interesting to note that, after identifying $k = i\beta\,\Omega\,\pi/\ell$, the one-loop contribution reproduces the partition function for open rotating strings \eqref{dosrot} up to the exponential factor of $e^{{a}/{\beta}}$.
Thus, the modulation of the partition function due to non-vanishing angular potential $\Omega$ is exactly reproduced by the one-loop factor caused by imposing a non-vanishing area $J[X_i]$.

After inserting the evaluated path integral into \eqref{eq:APIRot2} one can perform the remaining integrals.
The integrals over the Lagrange multipliers $\mu_i$ are simple Gaussian integrals that yield 
\begin{align}
    d_{n,J}(x_i) &= \int\,d\beta \,e^{\beta\,n}\,\int\,dk\,e^{-i\,k\,J}\,\cN(\beta,k)\,\left(\frac{3\,\beta}{\pi^2\,\alpha'\,\ell}\right)^{\frac{c}{2}}\,\left(\frac{\beta}{\alpha'\,\ell^3}\right)^{\frac{c}{2}}\frac{1}{3\,\rb\left(\frac{\ell\,k}{\beta}\right)}\,\frac{\frac{\ell\,k}{\beta}}{\sinh\left(\frac{\ell\,k}{\beta}\right)}\,\nonumber\\*
    &\hspace{10em}\times \exp\left[-\frac{3\,\beta}{\pi\,\alpha'\,\ell}\left(\frac{x_1^2 + x_2^2}{3\,\rb\left(\frac{\ell\,k}{\beta}\right)}+x_a^2\right)\right]\,.
\end{align}

\paragraph{Determining the normalisation and the parameter length.}
To fix the normalisation factor $\cN(\beta, k)$ and the parameter length $\ell$, we impose that the density of states of strings with fixed $n$ and $J$, given by $d_{n,J}$, is equal to total number of random walk obtained by integrating the random walk density $d_{n,J}(x_i)$ over the entire space
\begin{align}
    \label{eq:AnJDensity}
    d_{n,J} \equiv  \int\,dx_i \,d_{n,J}(x_i)= \int\,d\beta \,e^{\beta\,n}\,\int\,dk\,e^{-i\,k\,J}\,\cN(\beta,k)\,\frac13\,\left(\frac{\beta}{\alpha'\,\ell^3}\right)^{\frac{c}{2}}\,\frac{\frac{\ell\,k}{\beta}}{\sinh\left(\frac{\ell\,k}{\beta}\right)}\,.
\end{align}
In going into the last expression we have assumed that one can exchange the order of integration and first evaluated the spatial Gaussian integrals.
The left-hand side needs to be provided by an independent method, for example by using operator methods \eqref{dos}.
Imposing \eqref{eq:AnJDensity} implies that the integrand of the above expression matches the partition function of rotating open strings in the thermodynamic limit%
\footnote{We have discussed the matching between the number of rotating  \emph{open} strings and  the number of \emph{closed} random walks in the Section~\ref{sec:randomwalk}.}
\begin{align}
\label{eq:AMatch1}
   Z(\beta,k) \propto \beta^{\frac{c}{2}}\,e^{\frac{a^2}{\beta}}\,\frac{\frac{\pi\,k}{\beta}}{\sinh\left(\frac{\pi\,k}{\beta}\right)} \equiv \cN(\beta,k)\,\beta^{\frac{c}{2}}\,\frac{\frac{\ell\,k}{\beta}}{\sinh\left(\frac{\ell\,k}{\beta}\right)} \,,
\end{align}
where we ignored all the unimportant constant numerical factors and used $a= \frac{\pi^2\,c}{6}$.
This identification then implies that $\ell = \pi$,%
\footnote{This value is also implicitly taken in equation (2.1) of \cite{Russo:1994ev}.} and that
\begin{align}
\label{eq:ANormalisation}
    \cN(\beta,k) = e^\frac{a}{\beta}\,,
\end{align}
up to irrelevant constant prefactors.
Once this normalisation is taken into account, the density of states integral \eqref{eq:AnJDensity} becomes identical to \eqref{eq:denstates} discussed in Section~\ref{sec:RotatingStringSize} and can be evaluated to  \eqref{dos}.

Let us note that the normalisation factor \eqref{eq:ANormalisation} is $k$-independent, which is the consequence of evaluating the path integral by including the one-loop factor \eqref{eq:AOneLopp}.
One can use this $k$-independence to simplify the external input needed to make the random walk work.
Namely, by integrating \eqref{eq:AnJDensity} over all possible values of $J$ one effectively evaluates the partition function at $k=0$
\begin{align}
\label{eq:Adndef}
     \int\,dJ\, d_{n,J} = \int\,d\beta \,e^{\beta\,n}\,\cN(\beta,k=0)\,\frac13\,\left(\frac{\beta}{\alpha'\,\ell^3}\right)^{\frac{c}{2}}\equiv d_{n}\,,
\end{align}
which is then equal to the number of strings states at level $n$ with all possible values of angular momentum, given in \eqref{dnTM}.
Again, this is equivalent to identifying the argument of the above integral with the partition function of static strings in the thermodynamic limit \eqref{Zbetathermo}
\begin{align}
\label{eq:PartfunIden}
    \cN(\beta,k=0)\,\frac13\,\left(\frac{\beta}{\alpha'\,\ell^3}\right)^{\frac{c}{2}} \propto \beta^{\frac{c}{2}}\,e^{\frac{a}{\beta}}\,,
\end{align}
Ignoring the numerical prefactors, this again leads to \eqref{eq:ANormalisation}. 
If one assumes that the complete $k$-dependence comes from the one-loop contribution, one obtains the same result as if one imposes \eqref{eq:AnJDensity}.

To summarise, we find that the number of paths of squared length $n$ and angular momentum $J$ that go through point $x_i$ is given by
\begin{equation}
\label{eq:ARWDensity}
    d_{n,J}(x_i) = \int d\beta \,e^{\beta\,n}\int\,dk\,e^{-ikJ}e^{\frac{a}{\beta}}\,\beta^{\frac{c}{2}}\,\left(\frac{3\,\beta}{\pi^3\,\alpha'}\right)^{\frac{c}{2}}\!\!\frac{1}{3\rb\left(\frac{\pi\,k}{\beta}\right)}\,\frac{\frac{\pi\,k}{\beta}}{\sinh\left(\frac{\pi\,k}{\beta}\right)} \exp\left[-\frac{3\,\beta}{\pi^2\,\alpha'}\left(\frac{x_1^2 + x_2^2}{3\,\rb\left(\frac{\pi\,k}{\beta}\right)}+x_a^2\right)\right],
\end{equation}
where we have ignored some overall constant prefactors. This is the result quoted in the main text in equation~\eqref{eq:RWDensity}.
The spatial dependence is a normal distribution whose width depends on the temperature and, for the directions in the plane of rotation, also on the angular potential. 
We discuss this in more detail in Appendix~\ref{sapp:CanEns}.

\subsection*{Path integral for the closed string}
\label{sapp:Closed}

Up to this point, we have compared a single \emph{closed} random walk with fixed length square $n$ and area $J$ to \emph{open} bosonic strings with  fixed level $n$ and angular momentum $J$. 
We have found perfect agreement between the sizes, which may be surprising given the mismatch between the boundary conditions imposed on the random walk and on the open strings.
However, as already discussed in Section~\ref{sec:randomwalk}, this is just the consequence of looking at only one of the two sectors of the closed string.
In this subsection, we show that when we consider two independent random walks at the same temperature $\beta$, we can reproduce the partition function and the size of closed strings, as discussed in Appendix~\ref{sec:closed}.

Recall that closed string solutions can be written as a sum of left and right movers \eqref{eq:ClosedStringSum} and that the closed string partition function can be written as a product of two open string partition functions at equal temperature \eqref{eq:CLO2}, with the level matching condition imposed.
To set up the path integral, we use the heuristic interpretation \eqref{eq:Translation} and assign an independent random walk to each of the two sectors of the bosonic string.
We denote the two random walks by $X^{L}_i(s)$ and $X^{R}_i(s)$ and impose periodic boundary conditions on both
\begin{align}
    & X_i^{L}(0) = X_i^{L}(\pi) = x_i\,, && X_i^{R}(0) = X_i^{R}(\pi) = x_i\,.
\end{align}
In other words, we are interested in a pair of random walks which both start and end at the same point.
In addition, since the position is given by the sum of left and right movers, we need to modify the center of mass condition \eqref{eq:AvgXdef} to 
\begin{align}
\label{eq:ACOMclosed}
    \avg{X^L_i +X^R_i}_{n,J} =0\,.
\end{align}
We want to count the number of pairs of closed random walks whose length squared is $n/2$,%
\footnote{We choose the length squared to be equal to $n/2$ instead of $n$ to match the number $n$ in \eqref{eq:ClosedNumber1}. This choice does not affect the scaling at large $n$.}
whose angular momenta are given by $J_L$ and $J_R$, and whose combined center of mass is at the origin. 
As in \eqref{eq:Density2}, we can express this as
\begin{equation}
    \label{eq:AClosedDensity}
    d_{n,J_L, J_R}(x_i)\! \propto\! \sum_{P^L_c(x_i)}\!\delta\left(\tfrac{n}{2}- n[X_i^L]\right)\,\delta(J_L - J[X_i^L])\!\sum_{P^R_c(x_i)}\!\delta\left(\tfrac{n}{2}- n[X_i^R]\right)\,\delta(J_R - J[X_i^R])\,\delta(\avg{X^L_i +X^R_i})\Bigg|_{\beta_L = \beta_R}\!\!,
\end{equation}
where we already imposed that the two Lagrange multipliers enforcing the condition on the length squared are equal.
Using the same logic as in the first subsection of this appendix, we can write this sum as a pair of Euclidean path integrals
\begin{align}
    \label{eq:APIRotClosed}
    d_{n,J_L, J_R}(x_i)&= \int\,d\beta \,e^{\beta\,n} \int dk_L\,dk_R\, e^{- i\,k_L\,J_L}\,e^{-i\,k_R\,J_R}\,\cN(\beta, k_L,k_R) \int\,d\mu_i\nonumber\\*
    &\hspace{10em}\times \int_{X_i^L(0) = x_i}^{X_i^L(\pi)= x_i}\,\cD X_i^L\,e^{-I_{k_L}[X_i^L]}\,\int_{X_i^R(0) = x_i}^{X_i^R(\pi)= x_i}\,\cD X_i^R\,e^{-I_{k_R}[X_i^R]}\,,
\end{align}
where we  assumed that one can exchange the order of integration and first evaluate the the path integrals.
The action $I_k[X_i]$ is given in \eqref{eq:Action1}, and we again introduced a normalisaton constant $\cN(\beta, k_L,k_R)$, which will be fixed so that the total number of random walks, obtained by integrating the above expression over the entire space, reproduces the partition function of the \emph{closed} string.

The two path integrals can be evaluated as before, with 
\begin{align}
        \int_{X_i^L(0) = x_i}^{X_i^L(\pi)= x_i}\,\cD X_i^L\,e^{-I_{k_L}[X_i^L]} \propto \beta^{\frac{c}{2}}\frac{\frac{\pi\,k_L}{\beta}}{\sinh\left(\frac{\pi\,k_L}{\beta}\right)}\,e^{-I_{k_L}^{\rm (c)}}\,, 
\end{align}
and an analogous expression holds for the $X_i^R$ path integral.
We have ignored irrelevant constant prefactors. The classical on-shell action is given by  \eqref{eq:AActionRot} with $\ell = \pi$.
The total on-shell action is then just the sum of the left and right-moving parts
\begin{align}
\label{eq:AActionClosed}
    I_{k_L + k_R}^{\rm (c)} &= I_{k_L}^{\rm (c)}+ I_{k_R}^{\rm (c)} \\*
    &= 2\left[\frac{3\,\beta}{\pi^2\,\alpha'}\left(\frac{x_1^2 + x_2^2}{3\, \rb_{\rm LR}}+x_a^2\right)
    + \frac{\pi^4\,
    \alpha'\,\rb_{\rm LR}}{4\,\beta}\sum_{j=1,2}\left(\mu_j+ \frac{2\,i\,\beta\,x_j}{\pi^3\,\alpha'\,\rb_{\rm LR}} \right)^2 
    +\frac{\pi^4\,\alpha'}{12\,\beta}\left(\mu_a+ \frac{6\,i\,\beta\,x_a}{\pi^3\,\alpha'}\right)^2 \right],\nonumber
\end{align}
where 
\begin{align}
\label{eq:AAverageRho}
    \rb_{\rm LR}\equiv \frac{\rb\left(\frac{\pi\,k_L}{\beta}\right)+ \rb\left(\frac{\pi\,k_R}{\beta}\right)}{2}\,,
\end{align}
The action is in fact just twice the individual action \eqref{eq:AActionRot} with $\rb(\pi\,k/\beta)$ replaced by $\rb_{\rm LR}$, the average of the left and right moving contributions.
Inserting the evaluated path integrals into the sum \eqref{eq:APIRotClosed} and evaluating the $\mu_i$ integrals gives
\begin{align}
    \label{eq:APIRotClosed2}
    d_{n,J_L, J_R}(x_i)&= \int\,d\beta \,e^{\beta\,n} \int dk_L\,dk_R\, e^{- i\,k_L\,J_L}\,e^{-i\,k_R\,J_R}\,\cN(\beta, k_L,k_R) \,\beta^{c}\frac{\frac{\pi\,k_L}{\beta}}{\sinh\left(\frac{\pi\,k_L}{\beta}\right)}\,\frac{\frac{\pi\,k_R}{\beta}}{\sinh\left(\frac{\pi\,k_R}{\beta}\right)}\nonumber\\*
    &\quad \times \left(\frac{6\,\beta}{\pi^3\alpha'}\right)^{\frac{c}{2}}\,\frac{1}{3\,\rb_{\rm LR}}\,\exp\left[- \frac{6\beta}{\pi^2\,\alpha'}\left(\frac{x_1^2 + x_2^2}{3\, \rb_{\rm LR}}+x_a^2\right)\right]\,.
\end{align}
To find $\cN(\beta, k_L,k_R)$, we again demand that the integral of this expression over the entire space reproduces the total number of closed strings states of level $n$ and fixed angular momentum $J_L$ and $J_R$, as given in \eqref{eq:ClosedNumber2}
\begin{align}
\label{eq:AidentificationClosed}
    \int\,dx\,d_{n,J_L, J_R}(x_i) \equiv d_{n, J_L, J_R}\,.
\end{align}
This means that we are equating the number of states of a closed string to the total number of pairs of random walks of fixed $J_L$ and $J_R$ and fixed length squared $n$.
Assuming that one can exchange the order of integration, one finds 
\begin{equation}
\label{eq:AdxIntegralClosed}
      \int \!dx_i\,d_{n,J_L, J_R}(x_i) \!= \! \int\! d\beta \,e^{\beta\,n} \!\int dk_L\,dk_R\, e^{- i\,k_L\,J_L}e^{-i\,k_R\,J_R}\cN(\beta, k_L,k_R) \,\beta^{c}\,\frac{\frac{\pi\,k_L}{\beta}}{\sinh\left(\frac{\pi\,k_L}{\beta}\right)}\,\frac{\frac{\pi\,k_R}{\beta}}{\sinh\left(\frac{\pi\,k_R}{\beta}\right)}\,.
\end{equation}
Comparing this expression to \eqref{eq:ClosedNumber1}, we find that \eqref{eq:AidentificationClosed} holds if, up to constant factors,
\begin{align}
    \cN(\beta, k_L,k_R) = \cN(\beta, k_L)\,\cN(\beta, k_R)\equiv  e^{\frac{a_L}{\beta}}\,e^{\frac{a_R} {\beta}}\,.
\end{align}
We interpret this normalisation in the same way as in the case of a single random walk: In both the left and right moving sector we need to impose an effective cut-off due to the fact that we are approximating an inherently quantum system with continuous curves.

All in all, the number of pairs of closed random walks with fixed $n$ and $J_L$ and $J_R$ going through point $x_i$ is given by
\begin{align}
    \label{eq:AClosedResFinal}
    d_{n,J_L, J_R}(x_i)&= \int\,d\beta \,e^{\beta\,n} \int dk_L\,dk_R\, e^{- i\,k_L\,J_L}\,e^{-i\,k_R\,J_R}\beta^{c}\,e^{\frac{a_L+a_R}{\beta}}\frac{\frac{\pi\,k_L}{\beta}}{\sinh\left(\frac{\pi\,k_L}{\beta}\right)}\,\frac{\frac{\pi\,k_R}{\beta}}{\sinh\left(\frac{\pi\,k_R}{\beta}\right)}\nonumber\\*
    &\quad \times \left(\frac{6\,\beta}{\pi^3\alpha'}\right)^{\frac{c}{2}}\,\frac{1}{3\,\rb_{\rm LR}}\,\exp\left[- \frac{6\beta}{\pi^2\,\alpha'}\left(\frac{x_1^2 + x_2^2}{3\, \rb_{\rm LR}}+x_a^2\right)\right]\,.
\end{align}

\paragraph{Size transverse to the rotation plane.}

Using the above random walk model one can reproduce the size of highly excited closed string for arbitrary values of $J_L$ and $J_R$ given in Appendix~\ref{sec:closed}. 
We calculate the average size by using 
\begin{align}
    \avg{A}_{n,J_L, J_R} \equiv \frac{1}{d_{n,J_L, J_R}}\int\,dx_i\,A\,d_{n, J_L.J_R}(x_i)\,.
\end{align}
The size orthogonal to the plane of rotation is given by 
\begin{align}
    \avg{x_{\perp}^2}_{n,J_L,J_R}= \frac{\pi^2\,\alpha'}{12\,d_{n,J_L,J_R}}\int\,d\beta \,e^{\beta\,n}\! \int dk_L\,dk_R\, e^{- i\,k_L\,J_L}\,e^{-i\,k_R\,J_R}\beta^{c-1}\,e^{\frac{a_L+a_R}{\beta}}\frac{\frac{\pi\,k_L}{\beta}}{\sinh\left(\frac{\pi\,k_L}{\beta}\right)}\,\frac{\frac{\pi\,k_R}{\beta}}{\sinh\left(\frac{\pi\,k_R}{\beta}\right)}\,,
\end{align}
The integral transforms in the above expression are exactly the same as in the evaluation of the closed string partition function \eqref{eq:AdxIntegralClosed}, only with the exponent of $\beta$ being shifted $c\to c-1$.
The argument of the integral transforms is then exactly \eqref{eq:ClosedPerpInt} in the thermodynamic limit: the expression that is obtained using operator method. 
Thus evaluating the average size of the random walk yields \eqref{eq:PerpSizeClosed}, in agreement with the operator methods.

\paragraph{Size along the plane of rotation.}

The average size in the plane of rotation 
\begin{align}
    \label{eq:AParaSizeClosed}
    &\avg{x_{\parallel}^2}_{n,J_L,J_R} = \frac{1}{d_{n,J_L,J_R}}\,\int\,dx_i\,x_{\parallel}^2\,d_{n,J_L,J_R}(x_i)\\*
    &= \frac{\pi^2\,\alpha'}{8\,d_{n,J_L,J_R}}\left[ \int\,d\beta \,e^{\beta\,n}\beta^{c-1}\,e^{\frac{a_L+a_R}{\beta}}\,\! \int dk_L\, e^{- i\,k_L\,J_L}\,\rb\left(\frac{\pi\,k_L}{\beta}\right)\,\frac{\frac{\pi\,k_L}{\beta}}{\sinh\left(\frac{\pi\,k_L}{\beta}\right)}\int \,dk_R\,e^{-i\,k_R\,J_R}\,\,\frac{\frac{\pi\,k_R}{\beta}}{\sinh\left(\frac{\pi\,k_R}{\beta}\right)}\right.\nonumber\\*
    &\quad +\left. \int\,d\beta \,e^{\beta\,n}\beta^{c-1}\,e^{\frac{a_L+a_R}{\beta}}\,\! \int dk_L\, e^{- i\,k_L\,J_L}\frac{\frac{\pi\,k_L}{\beta}}{\sinh\left(\frac{\pi\,k_L}{\beta}\right)}\int \,dk_R\,e^{-i\,k_R\,J_R}\,\rb\left(\frac{\pi\,k_R}{\beta}\right)\,\frac{\frac{\pi\,k_R}{\beta}}{\sinh\left(\frac{\pi\,k_R}{\beta}\right)} \right]\nonumber\,,
\end{align}
contains two terms that each include two Fourier transforms which we have already encountered in the open-string evaluation.
It is then straigthforward to show that
\begin{align}
     \label{eq:AParaSizeInt2}
    &\avg{x_{\parallel}^2}_{n,J_L,J_R} =\frac{\pi^2\,\alpha'}{8\,d_{n,J_L,J_R}} \int\,d\beta \,e^{\beta\,n}\beta^{c+1}\,e^{\frac{a_L+a_R}{\beta}} \left(\frac{\log{\left(1 + e^{-\beta\,|J_L|}\right)} + \frac{\beta\,|J_L|}{1 + e^{\beta\,|J_L|}}}{\cosh^2\left(\frac{\beta \,|J_R|}{2}\right)} + J_L \leftrightarrow J_R\right)\,.
\end{align}
Setting $a_L = a_R = a$ and recalling that $b= c/2$, we see that the argument of the inverse Laplace transform in the above equation is exactly given by \eqref{eq:InvLapArgument} obtained by operator methods.
Therefore, evaluating the above integral using the saddle point approximation will give an exact match between the size of the random walks and the rotating closed string for all values of $J_L$ and $J_R$, as given in Appendix~\ref{sec:closed}, 

\section{Sizes expressed in the \texorpdfstring{$\beta$}{beta} and \texorpdfstring{$\Omega$}{Omega} variables}
\label{sapp:CanEns}

In the main text we showed that using the spatial distribution \eqref{eq:ARWDensity} of the random walk one reproduces the sizes of highly excited rotating strings for all values of $J$.
Finding the explicit expressions for the sizes involves evaluating a Fourier and Laplace transform, which may become tedious if higher moments of the distribution are evaluated.%
\footnote{This may be of interest when evaluating multipole moments of rotating strings, for example.}
In this appendix, we take an alternative approach of calculating the sizes using $\beta$ and $\Omega=-i\,
k/\beta$ 
and show how to relate them to $n$ and $J$ bypassing the evaluation of the aforementioned integrals using thermodynamic relations between these variables.

When determining the normalisation factor one effectively imposes that the integral of the distribution over the entire space, \emph{before} one performs the Fourier and Laplace transforms, is equal to the partition function of rotating strings \eqref{eq:AMatch1}.
It is then natural to take a step back and identify the integrand of \eqref{eq:ARWDensity} with the  spatial distribution of random walks at fixed (inverse) temperature $\beta$ and angular potential $k$
\begin{align}
    \label{eq:ACanonicalEnsDist}
    Z(\beta,k;x_i) \equiv e^{\frac{a}{\beta}}\,\beta^{\frac{c}{2}}\,\left(\frac{3\,\beta}{\pi^3\,\alpha'}\right)^{\frac{c}{2}}\!\!\frac{1}{3\rb\left(\frac{\pi\,k}{\beta}\right)}\,\frac{\frac{\pi\,k}{\beta}}{\sinh\left(\frac{\pi\,k}{\beta}\right)} \exp\left[-\frac{3\,\beta}{\pi^2\,\alpha'}\left(\frac{x_1^2 + x_2^2}{3\,\rb\left(\frac{\pi\,k}{\beta}\right)}+x_a^2\right)\right]\,.
\end{align}
By construction, this distribution is already normalised so that when integrated over the entire space, it reproduces the partition function of the open string \eqref{dosrot}.

Let us now use the distribution in the $\beta$ and $k$ variables to calculate the sizes of random walks. Similar to before, we define the expected value of an observable $A$ as
\begin{align}
\label{eq:ABetaAnsDef}
    \avg{A}_{\beta,k} \equiv \frac{1}{Z(\beta,k)}\,\int\,dx_i\,A\,Z(\beta,k;x_i)\,,
\end{align}
where by construction $Z(\beta,k) = \int\,dx_i\,Z(\beta,k;x_i)$.
We find that the expected values for the size of the random walk are given by
    \begin{align}
    \label{eq:ASizebetak}
        &\avg{x_{\perp}^2}_{\beta,k} = \frac{\pi^2\,\alpha'}{6\beta}\,,  && \avg{x_{\parallel}^2}_{\beta,k} = \frac{\pi^2\,\alpha'}{2\beta}\,\rb\left(\frac{\pi\,k}{\beta}\right)\,,
    \end{align}
which indeed matches the sizes found using the operator methods in \eqref{perpsize} and \eqref{parallsize}, when $Z(\beta,\Omega)$ is divided out.    
Since these expressions also determine the width of the Gaussian distribution in \eqref{eq:RWDensity}, it may be worthwhile to look at them in more detail.
The size of the random walk in the directions orthogonal to the plane of rotation does not depend on the angular potential, but scales linearly with the temperature in string units
\begin{align}
\label{eq:APerpSizeScaling}
    \avg{x_{\perp}^2}_{\beta,k} \sim \frac{\alpha'}{\beta} \sim \frac{T}{T_H}\,\ell_s\,,
\end{align}
where we used that the Hagedorn temperature $T_H$ is inversely proportional to the string length.
Such a scaling with temperature is intuitive, since larger thermal fluctuations allow the random walk to explore a larger space.
While this also suggests that the size vanishes at zero temperature, we must recall that this lies beyond the regime of validity of  the random walk model.
To analyse the expression for the size in the plane of rotation, recall that $k = i\beta\,\Omega$, which gives
\begin{align}
\label{eq:AParaSizebeta}
    \avg{x_{\parallel}^2}_{\beta,\Omega} = \frac{\pi^2\,\alpha'}{2\beta}\,\rho\left(\pi\,\Omega\right)\,, 
\end{align}
where
\begin{align}
    \label{eq:AEuclideanRhotation}
    \rho(\pi\,\Omega) = \frac{1-\pi\,\Omega\,\cot (\pi\,\Omega)}{\pi^2\,\Omega^2}\,,
\end{align}
is, up to an overall factor of $\pi^2$,  the same as the rhotation function  \eqref{rhohightemp} found in the high-temperature limit.
This change is important because now \eqref{eq:AParaSizebeta} separates into a term that is linearly dependent on the temperature (for the same reason as in \eqref{eq:APerpSizeScaling}) multiplied by a temperature-independent term that contains all the dependence on the angular potential.
For small values of the angular potential ($\Omega \ll 1$) we find the leading order correction is positive and quadratic in $\Omega$,
\begin{align}
    \avg{x_{\parallel}^2}_{\beta,\Omega} \sim\left(1+ \frac{\pi^2}{15}\,\Omega^2+\ldots\right) \frac{T}{T_H}\,\ell_s\,.
\end{align}
In fact, for $|\Omega|<1$, the size in the plane of rotation increases when the magnitude of the angular potential is increased.
In the regime of large $\Omega$ the difference between $\rb(\pi\,\Omega)$ and $\rho(\pi\,\Omega)$ becomes crucial. While the former goes smoothly to 0 for large values of the argument, the latter diverges at integer multiples of $\pi$.
In particular
\begin{align}
\label{eq:Arbexp}
    \rho(\pi\,\Omega)\xrightarrow[\Omega\to \pm 1]{}\frac{1}{\pi^2\left(1\mp\Omega\right)}\,,
\end{align}
from which it follows that 
\begin{align}
     &\avg{x_{\parallel}^2}_{\beta,\Omega} \sim \frac{1}{\left(1\mp \Omega\right)}\frac{T}{T_H}\,\ell_s\,,&&\text{as}\quad \Omega \to \pm 1\,.
\end{align}
meaning that the size diverges as the angular potential approaches the critical values.

\paragraph{Difference between $(n,J)$ and $(\beta,\Omega)$ variables.}
Only the $x_i$-dependent exponential factor in \eqref{eq:ACanonicalEnsDist} is important when determining the sizes \eqref{eq:ASizebetak}.
All the remaining prefactors, including the $e^{\frac{a}{\beta}}$, which are crucial to determine the number of random walks and their sizes expressed in the $n$ and $J$ variables, do not need to be included to obtain the final result. 
In other words, the random walk in $\beta$ and $\Omega$ variables doesn't require any ``external'' knowledge of the partition function to reproduce the sizes of rotating strings.
This is because one can normalise the distribution \eqref{eq:ACanonicalEnsDist} such that its spatial integral is equal to 1 and treat it as a probability distribution, as in \cite{Manes:2004nd}.
Working with a normalised probability distribution implicitly takes care of any issues with the infinite number of paths passing through a point, as in this case one  does not ``count'' the number of paths passing through a point $x_i$ (as in \eqref{eq:RWDensity}) but rather the fraction of paths passing in an infinitesimal volume of space around $x_i$.

The downside of working with a normalised probability distribution is that one cannot directly change the variables from $(\beta, \Omega)$ to $(n,J)$, since a probability distribution is ratio of two quantities which need to be transformed individually. 
This is just the consequence of the non-commutativity between the composition of functions and integral transforms.
For the exact same reason one cannot first calculate the sizes using the $\beta$ and $\Omega$ variables as in \eqref{eq:ASizebetak} and then integral transform these expressions: One has to start by transforming the distribution \eqref{eq:ACanonicalEnsDist} and only then calculate the sizes.

\paragraph{Using thermodynamic relations to find the sizes in $n$ and $J$ variables.}
One can avoid the integral transforms by utilising the thermodynamic relations between $(\beta,\Omega)$ and their conjugate variables, which is justified, since we are always working in the high-temperature limit.
We show that this reproduces the correct scaling for the sizes of strings in the regime when $J\sim\coo{1}$ and $J\sim \coo{n}$.%
\footnote{The calculation for typical strings where $J\sim \coo{\sqrt{n}}$ again involves inverting a transcendental equation, so we do not give the details here. However, the correct scaling for the string sizes with $n$ and $J$ is obtained for this regime as well.}

In the thermodynamic limit, we can write the string partition function as
\begin{align}
\label{eq:ALegendreDef}
    Z(\beta \approx 0,\Omega)\propto \beta^{\frac{c}{2}} e^{\frac{a}{ \beta}} \frac{\pi\,\Omega}{\sin(\pi \Omega)}\approx e^{-\beta (\bar n-\Omega\,\bar J)}\,,
\end{align}
where $\bar n$ and $\bar J$ are the expectation values of $n$ and $J$ expressed in terms of $\beta$ and $\Omega$
\begin{subequations}
    \label{eq:Aexpval1}
    \begin{align}
        \bar n &= - \pd_\beta\log Z(\beta, \Omega)\Big|_{\beta\,\Omega} = \frac{a}{\beta^2}- \frac{c}{2\,\beta}+ \frac{\pi^2\,\Omega^2}{\beta}\rho(\pi\,\Omega)\,,\\*
        \bar J &= \pd_{\beta\Omega}\log Z(\beta, \Omega)\Big|_{\beta} = \frac{\pi^2\,\Omega}{\beta}\,\rho(\pi\,\Omega)\,.\label{eq:ArhoJ}
    \end{align}
\end{subequations}
Note that these two expression can be combined to give 
\begin{align}
\label{eq:ATDExp1}
    \bar n - \Omega\,\bar J = \frac{a}{\beta^2}- \frac{c}{2\,\beta}\,.
\end{align}

\begin{figure}[t]
    \centering
    \includegraphics[width=0.9\linewidth]{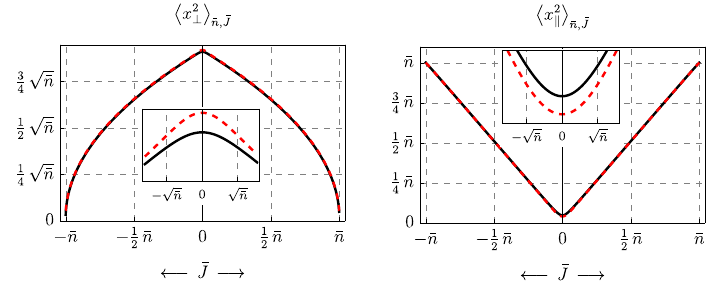}
    \caption{\small The size of strings with $\bar n=1000$ in $D=5$ orthogonal to the plane of rotation (left) and in the plane of rotation (right) obtained by using the thermodynamic relation \eqref{eq:ALegendreDef} (\textbf{black}) and integral transforms (dashed \textbf{\red{red}}) discussed in the main text. 
    We note that the two results are extremely close and differ only for small values of $\bar J$, where deviations are expected.
    In particular, the former, simpler, method captures the relevant scaling with $\bar n$ and $\bar J$ without the need to evaluate the complicated Fourier and Laplace transforms.
    }
    \label{fig:tdvsexact}
\end{figure}

The idea is to invert \eqref{eq:Aexpval1} and express  $\beta$ and $\omega$ in terms of $\bar n$ and $\bar J$. 
One can do this numerically and we show the plots for fixed $\bar n$ and see the dependence on $\bar{J}$ in Figure~\ref{fig:tdvsexact}, where we compare these results to the ones presented in the main text where we evaluated the two integral transforms.
In the regimes where $\bar J\sim \coo{1}$ and $\bar J\sim\coo{n}$ we can also derive some analytic results whose scaling matches the expressions quoted in \eqref{eq:ResPerp} and in \eqref{eq:ResParallel}.

\subparagraph{Sizes at small angular momentum ($\bar J\sim \coo{1}$).}
In this regime it is consistent to take $\Omega \ll \bar n^{-\frac12}$, where $\bar n \gg 1$, and expand $\rho(\pi\,\Omega)$ in small $\Omega$.
In the large $\bar n$ expansion one finds
\begin{align}
    \beta = \sqrt{\frac{a}{\bar n}}+ \ldots + \frac{c\,\bar J^2}{4\,\bar n^2}+ \ldots\,,&& \Omega = \frac{c\,\bar J}{2\,\sqrt{a\,\bar n}}+\ldots\,,
\end{align}
where we also omitted subleading $\bar J$-independent terms.
Using these expressions in~\eqref{eq:ASizebetak} one finds
\begin{subequations}
\label{eq:ATDSizesSmallJ}
    \begin{align}
        \avg{x_{\perp}^2}_{\bar n,\bar J} &= \frac{\pi^2\,\alpha'}{6}\left[\sqrt{\frac{\bar n}{a}} +\Gamma_1 + \frac{\Gamma_2}{\sqrt{\bar n}} +\coo{n^{-1}}
        \right]\,,\\*   
        \avg{x_{\parallel} ^2}_{\bar n,\bar J}&= \frac{\pi^2\,\alpha'}{6}\left[\sqrt{\frac{\bar n}{a}} +\Gamma_1 + \frac{\Gamma_2}{\sqrt{\bar n}}+ \frac{3\,\sqrt{a}}{5\,\pi^2}\,\frac{\bar J^2}{\sqrt{\bar n}}+\coo{n^{-1}}\right]\,,
    \end{align}
\end{subequations}
where
\begin{align}
\label{eq:TDconsts}
    \Gamma_1 = \frac{b}{2\,a}\,,\qquad \Gamma_2 = \frac{b^2}{8\,a^{\frac32}}\,.
\end{align}
Finally, the ratio between the sizes is then given by
\begin{align}
\label{eq:TDratio}
    \frac{\avg{x_{\perp}^2}_{\bar n,\bar J}}{\avg{x_{\parallel} ^2}_{\bar n,\bar J}}= 1- \frac{3\,a\,J^2}{5\,\pi^2\,\bar n}+ \coo{n^{-\frac32}}\,.
\end{align}
Comparing these expressions to the ones obtained using integral transforms \eqref{rperpsmallJ}-\eqref{eq:consts}, 
we observe that \eqref{eq:ATDSizesSmallJ} captures the right leading scaling with $\bar n$ and $\bar J$, but does not completely reproduce the numerical coefficients. 
This is somewhat expected, since this method is equivalent to evaluating both the Laplace and Fourier transform  using the saddle point method and ignoring any subleading corrections, including the one-loop factors. 
However these corrections are crucial for obtaining the right numerical factors for small $J$ in \eqref{eq:ResPerp} and \eqref{eq:ResParallel}, which explains the difference in numerical coefficients.%
\footnote{Interestingly, the coefficient multiplying the $J^2$ term in \eqref{eq:ATDSizesSmallJ} is very close to $\gamma_{\parallel}$ appearing in \eqref{eq:ResParallel}, with $3/(5\pi^2) \approx 0.0608$ and $\gamma_{\parallel} = (2\log2-1)/(8\log2) \approx 0.0697$. }
This can also be seen in the inset in Figure~\ref{fig:tdvsexact}.

\begin{figure}[t]
    \centering
    \includegraphics[width=0.9\linewidth]{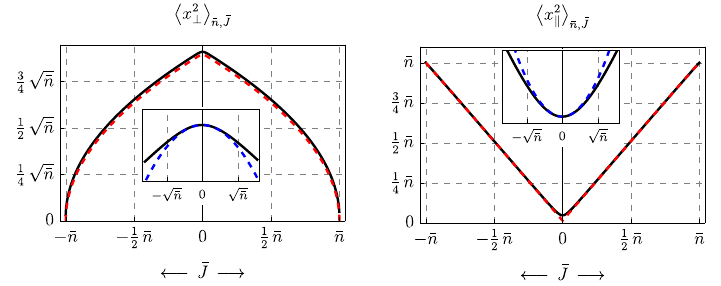}
    \caption{\small Size of strings in $D=5$ in directions orthogonal (left) and in the plane of rotation (right) as a function of $\bar{J}$ at fixed $\bar n$ with the inset plots zooming in on the region near the origin. 
    In black we plot the result of numerically inverting \eqref{eq:Aexpval1} using $\bar n=1000$.  In dashed \textbf{\red{red}} we plot the analytical results for $\bar J \sim \coo{\bar n}$, given  in \eqref{eq:ATDSizesSmallJ}.
    In dashed \textbf{\blue{blue}} we give the leading corrections at $\bar J \sim \coo{\bar 1}$.
    The leading behaviour is indeed quadratic in $\bar J$, but this approximation is not sufficient to capture the behaviour at $\bar J \sim \coo{\sqrt{\bar n}}$.
    }
    \label{fig:ASize}
\end{figure}
\subparagraph{Sizes at large angular momentum ($\bar J\sim \coo{\bar n}$).}
The scaling $|\bar J|\sim \coo{\bar n}$ is achieved by sending $\Omega\to \pm 1$, in which case we expand $\rho(\pi\,\Omega)$ around the nearest poles to the origin, as in \eqref{eq:Arbexp}, to get
\begin{align}
    \bar J = \pm\frac{1}{\beta(1\mp \Omega)}\,,\qquad \Longleftrightarrow\qquad |\bar J| = \frac{1}{\beta(1-|\Omega|)}
\end{align}
In this scaling regime, \eqref{eq:ATDExp1} can be inverted to give
\begin{align}
    \beta = \sqrt{\frac{a}{\bar n - |\bar J|}}\,,
\end{align}
where we assume $\bar n - |\bar J|$ is large and of order of $\bar n$.
Using this value for the inverse temperature, one finds that the sizes of the random walk orthogonal and in the plane  of rotation are given by
    \begin{align}
    \label{eq:ASizeLargeJTD}
    &\avg{x_{\perp}^2}_{\bar n,\bar J} = \frac{\pi^2\,\alpha'}{6}\,\sqrt{\frac{\bar n-|\bar J|}{a}}+ \ldots\,,&&
     \avg{x_{\parallel}^2}_{\bar n, \bar J} = \frac{\alpha'}{2}\,\frac{\bar J}{\Omega}=
     \frac{\alpha'}{2}\left( |\bar J| +\sqrt{\frac{\bar n-|\bar J|}{a}}\right) + \ldots\,,
\end{align}
where the ellipses denote subleading terms.
These expressions now exactly match the expressions obtained using the integral transforms given in \eqref{rperp} and \eqref{rparallel}, including the relevant numerical prefactors. 
As such, the ratio between the size in the direction orthogonal to the plane of rotation and the size in the plane of rotation is the same as in \eqref{eq:ResRatio} for $J\sim \coo{n}.$
These expressions are plotted in dashed red in Figure~\ref{fig:ASize} and we see that these asymptotic results nicely fit the numerical data for a large range of $\bar J$.

\bibliographystyle{JHEP}
\bibliography{companion_v2}

\end{document}